\documentclass[11pt,a4paper]{article}
\usepackage{jstyle}

\usepackage[utf8]{inputenc}
\usepackage{amsthm,amsmath,latexsym,amssymb,amsfonts,amssymb,amscd}
\usepackage{hyperref}
\usepackage{cancel}
\usepackage{color}
\usepackage{mathabx}
\usepackage{tikz}
\usetikzlibrary{matrix,backgrounds}

\usepackage[compat=1.1.0]{tikz-feynman}

\definecolor{rougef}{rgb}{0.56,0,0}
\definecolor{vertf}{rgb}{0,0.5,0}
\definecolor{bleuf}{rgb}{0,0,0.8}
\definecolor{violetf}{rgb}{0.5,0,0.5}

\newcommand{\Mi}{\text{Mi}}
\newcommand{\Mon}{\text{Mon}}

\usepackage[colorinlistoftodos,draft]{todonotes}

\title{\centering{Restrictions for $n$-Point Vertices in Higher-Spin Theories}}

\author[a,b]{Stefan Fredenhagen,}
\author[a]{Olaf Kr\"uger}
\author[c]{and Karapet Mkrtchyan}
\affiliation[a]{University of Vienna, Faculty of Physics, Boltzmanngasse 5, 1090 Vienna, Austria}
\affiliation[b]{Erwin Schr\"odinger International Institute for Mathematics and Physics, Boltzmanngasse 9, 1090 Vienna, Austria}
\affiliation[c]{Scuola Normale Superiore and INFN,
Piazza dei Cavalieri 7, 56126 Pisa, Italy}
\emailAdd{stefan.fredenhagen@univie.ac.at}
\emailAdd{olaf.krueger@univie.ac.at}
\emailAdd{karapet.mkrtchyan@sns.it}

\abstract{We give a simple classification of the independent $n$-point interaction vertices for bosonic higher-spin gauge fields in $d$-dimensional Minkowski spacetimes. We first give a characterisation of such vertices for large dimensions, $d \geq 2n - 1$, where one does not have to consider Schouten identities due to over-antisymmetrisation of spacetime indices. When the dimension is lowered, such identities have to be considered, but their appearance only leads to equivalences of large-$d$ vertices and does not lead to new types of vertices. We consider the case of low dimensions $(d<n)$ in detail, where a large number of Schouten identities leads to strong restrictions on independent vertices. 
We also comment on the generalisation of our results to the intermediate region $n \leq d \leq 2n - 2$. In all cases, the independent vertices are expressed in terms of elementary manifestly gauge-invariant quantities, suggesting that no deformations of the gauge transformations are induced.}

\begin{document}

\maketitle

\section{Introduction}
\label{sec:introduction}

In this paper, we investigate a Lagrangian formulation of higher-spin (HS) theories in arbitrary dimensions. 
The aim of this work is, in particular, to obtain restrictions for all possible independent interaction vertices of order $n\geq 4$ for massless higher-spin fields, extending the three-dimensional results of \cite{Fredenhagen:2019hvb}. Together with the earlier results on the cubic vertices \cite{Bengtsson:1986kh,Metsaev:1991mt,Metsaev:1991nb,Metsaev:2005ar,Manvelyan:2010jr,Conde:2016izb,Mkrtchyan:2017ixk,Kessel:2018ugi} (see also \cite{Sagnotti:2010at,Fotopoulos:2010ay,Manvelyan:2010je,Mkrtchyan:2011uh,Sagnotti:2013bha,Metsaev:2012uy}), this work intends to complete the classification of {\it all independent interacting deformations} of free massless HS Lagrangians \cite{Fronsdal:1978rb,Campoleoni:2012th} to the lowest order in the deformation parameters (coupling constants) in Minkowski spacetime of arbitrary dimensions $d\geq 3$.

HS Gravities \cite{Vasiliev:1990en,Prokushkin:1998bq,Vasiliev:2003ev} (see, e.g., \cite{bciv,Didenko:2014dwa} for reviews) are generalisations of Einstein's General Relativity which involve higher-spin gauge
fields. These are symmetric tensor (Fronsdal) fields\footnote{The index $s$ is the spin of the field and $\mu_i = 0, \ldots , d - 1$ in $d$ dimensions.} $\phi_{\m_1\dots\m_s}$, described by the Fronsdal
action \cite{Fronsdal:1978rb} at free level, describing massless particles of spin $s$ upon
quantisation.\footnote{In this paper, we will restrict ourselves to integer-spin (bosonic) fields for
simplicity.} A set of free HS fields can be described by a Lagrangian which is a sum of Fronsdal Lagrangians for spin $s$ fields. However, a full non-linear Lagrangian of \textit{interacting} Fronsdal fields is
not available to date.

Such theories are strongly constrained by gauge invariance, necessary for consistency. These gauge
transformations extend those of General Relativity -- spacetime reparametrisations, or diffeomorphisms -- to
larger symmetries, involving gauge parameters that are Lorentz tensors of rank $(s-1)$ for each massless spin
$s$ field. This extension of symmetries can potentially resolve some problems of General Relativity
(singularities, quantisation problem, etc.), making HS Gravity an attractive field of investigation.

The corresponding gauge transformation for free fields reads\footnote{The round (square) brackets denote (anti-)symmetrisation
  with weight one.}
\begin{equation}\label{freegauge}
  \delta^{(0)} \phi_{\mu_{1}\dots \mu_{s}}= s\, \partial_{(\mu_{1}}\epsilon_{\mu_{2}\dots \mu_{s})}\,,
\end{equation}
which generalises the well known expressions for massless vector fields ($s=1$) in gauge theory and the Graviton
($s = 2$) in linearised gravity theory. 

The naive intuition from lower-spin model building suggests that one can pick an arbitrary collection of fields,
including massless HS fields, and the gauge symmetries will partly constrain the interactions, leaving room for
a large parameter space of theories. It turns out that the severe constraints from HS gauge invariance rule out
theories with an arbitrary choice of the particle content. Therefore, one is easily led to negative results
if one chooses an arbitrary starting setup for constructing a theory with massless HS spectrum. This striking
difference from textbook examples makes it tempting to conclude (after some attempts) that such theories cannot
exist.

The problem can be traced to the global symmetries of the theory (see, e.g., \cite{Joung:2014qya}). One can
construct a HS theory by looking for suitable global symmetry algebras, which have to satisfy the so-called
admissibility condition \cite{Konshtein:1988yg}. In this way, infinitely many potential candidate algebras are
ruled out (see, e.g., \cite{Joung:2014aba,Joung:2019wwf}). This is crucial in deriving a list of admissible HS algebras
\cite{Konstein:1989ij,Vasiliev:1989re} and constructing full non-linear HS equations
\cite{Vasiliev:1990en,Prokushkin:1998bq,Vasiliev:2003ev} in the frame formulation, proving the existence of a
theory with massless HS fields. The resulting theory, however, has unusual properties:
there is an infinite tower of massless higher-spin fields with $s=0,1,2,\dots$ and a necessarily non-zero
cosmological constant \cite{Fradkin:1986qy}. The need for a non-zero cosmological constant is related to
diffeomorphism transformations, as explained in \cite{Fradkin:1987ks,Joung:2013nma}, essential for the
Fradkin-Vasiliev solution to the Aragone-Deser problem \cite{Aragone:1979hx}. This argument, together with the
holographic conjectures (see, e.g., \cite{Giombi:2016ejx,Gaberdiel:2012uj}) motivated the intense studies of HS
interactions, especially in (Anti-)de Sitter ($(A)dS$) background
\cite{Manvelyan:2009tf,Bekaert:2009ud,Mkrtchyan:2010zz,Vasilev:2011xf,Joung:2011ww,Joung:2012fv,Manvelyan:2012ww,Boulanger:2012dx,Francia:2016weg,Sleight:2017fpc,Bekaert:2014cea,Sleight:2016dba,Karapetyan:2019psg}.

The frame-like formulation of HS gravities that led to successful developments (including Vasiliev's non-linear
equations \cite{Vasiliev:1990en,Prokushkin:1998bq,Vasiliev:2003ev} and their recent generalisations \cite{Grigoriev:2018wrx,Sharapov:2019vyd}) registered less progress so far in understanding the corresponding Lagrangian formulation. On the other hand, the metric-like formulation \cite{Fronsdal:1978rb,Campoleoni:2012th,Francia:2016weg} is a simple suitable setup for classifying interaction
vertices and deriving restrictions on interacting Lagrangians. Here, we work in the framework of the
Noether-Fronsdal program (see \cite{Bengtsson:1983pd,Berends:1984rq,Bekaert:2006us,fms1,Fotopoulos:2008ka,Zinoviev:2008ck,Boulanger:2008tg,Manvelyan:2009vy,Manvelyan:2010wp,Taronna:2011kt,Ruehl:2011tk,Bengtsson:2016hss,Taronna:2017wbx,Roiban:2017iqg,Ponomarev:2017qab} for related literature and \cite{Kessel:2018ugi} for a recent summary of the
status of the problem) to classify independent vertices of order $n\geq 4$ in arbitrary dimensions $d\geq 3$,
generalising the $d=3$ results obtained earlier in \cite{Fredenhagen:2019hvb}.

The situation is different only in three dimensions, where the interacting HS theories can admit arbitrary
Einstein backgrounds (including Minkowski) as well as a finite spectrum of massless HS fields (see, e.g., \cite{Campoleoni:2010zq,Campoleoni:2012hp,Fredenhagen:2014oua,Gwak:2015vfb,Campoleoni:2017xyl}). However, such
massless HS fields do not correspond to propagating particles in $d=3$, while the inclusion of matter leads to a
situation similar to the higher-dimensional story in many ways.

It is also interesting to note here, that four dimensions is also somewhat special: there, the Aragone-Deser problem is relevant at the Lagrangian level for descriptions using Fronsdal fields, while the minimal coupling to gravity is available at the level of amplitudes (see, e.g., \cite{Benincasa:2011pg,Conde:2016izb,Ponomarev:2016cwi,Nagaraj:2018nxq})
and light-cone vertices \cite{Bengtsson:1986kh,Metsaev:1991mt,Metsaev:1991nb,Ponomarev:2016lrm,Metsaev:2018xip,Metsaev:2019aig,Ponomarev:2017nrr} together with other vertices that are absent in spacetime dimensions $d\geq 5$. Most importantly, these vertices seem to be essential for the consistency of the non-linear theory \cite{Metsaev:1991mt,Metsaev:1991nb} in four dimensions.

The Noether-Fronsdal program is a systematic approach to perturbatively construct a Lagrangian $\mathcal{L}$ for
an arbitrary interacting HS theory order by order. In this procedure, $\mathcal{L}$ is expanded in powers of
small parameters $g_n$,
\begin{equation}
  \label{eq:noether-exp}
  \mathcal{L}=\mathcal{L}_{2} + \sum_{n\geq 3} g_{n} \mathcal{L}_n + O(g_{n}^2)\,.
\end{equation}
Here, $\mathcal{L}_2$ denotes the free Fronsdal Lagrangian and another sum over the different kinds of $n$-point
vertices $\mathcal{L}_n$ is suppressed. 

The action must be gauge-invariant, hence, $\delta \mathcal{L}$ equals a total derivative, where $\delta$ is
obtained by a deformation of the free gauge transformation $\delta^{(0)}$,
\begin{equation*}
  \delta=\delta^{(0)}+ \sum_{k \geq 1} \delta^{(k)}\,.
\end{equation*}
Here, the deformation $\delta^{(k)}$ is of $k$-th order in the fields. Since our aim is to find constraints for
the \textit{independent} vertex structures (i.e.\ linear in the coupling constants\footnote{Gauge invariance
  provides constraints to fix the terms proportional to higher powers of coupling constants. We are interested
  here in the structures that parametrise the non-trivial deformations at the lowest order in the coupling
  constants.}), the $n$-point vertex must satisfy
\begin{equation}\label{NE}
  \delta^{(0)}\mathcal{L}_{n} + \delta^{(n-2)} \mathcal{L}_{2} = 0\ \ \text{up to total derivatives}\,.
\end{equation}

In this paper, we find restrictions for all independent $n$-point vertices $\mathcal{L}_n$ for massless HS
fields in arbitrary dimension $d \geq 3$, such that they satisfy Eq.~(\ref{NE}).\footnote{We consider a flat
Minkowski spacetime but comment also on $(A)dS$ backgrounds in Section \ref{sec:discussion}.} From that, we
deduce a simple classification of vertices. For a summary of the explicit results, see the beginning of
Section~\ref{sec:discussion}.

The paper is organised as follows: In Section~\ref{sec:preliminaries}, we set up notations and provide the mathematical framework for
our analysis. There, we discuss that we have to analyse three different cases separately: Large dimensions
$d \geq 2n - 1$ (see Section~\ref{sec:case-2n-leq}), low dimensions $d < n$ (see Section~\ref{sec:case-n-}) and
the intermediate case (see Section~\ref{sec:lowerd} and comments in Section~\ref{sec:discussion}). We mostly consider parity-even vertices,
but give a generalisation to parity-odd vertices in Section~\ref{sec:parity-odd-vertices}. We finally conclude
in Section~\ref{sec:discussion}.

\section{Preliminaries}
\label{sec:preliminaries}

We want to constrain the $n$-point independent vertices $\mathcal{L}_n$ that may constitute the lowest order deformations of the free Lagrangian for massless HS fields. For this purpose, we restrict ourselves to the traceless and transverse (TT) sector of the Lagrangian as in \cite{Fredenhagen:2019hvb}. We refer the reader to a more detailed discussion\footnote{For massive fields, the restriction to TT is a choice of field redefinition freedom and therefore completely legitimate (as discussed in Appendix B.1 of \cite{Joung:2019wwf}), while for massless fields it can introduce subtleties in some formulations, as discussed in Section 4.5 of \cite{Francia:2016weg}.} on the TT vs off-shell vertices in \cite{Francia:2016weg}.
In the following, we therefore
assume that the tensors $\phi_{\mu_1\dots\mu_s}$ that describe the gauge fields, are traceless, divergence-free and
the corresponding free equation of motion is given by the (massless) Klein-Gordon equation, hence
\begin{equation}
  \label{eq:phi-trace-div}
  g^{\mu_1 \mu_2} \phi_{\mu_1\dots\mu_s} = 0\,, \qquad \partial^{\mu_1} \phi_{\mu_1 \cdots \mu_s} = 0\,, \qquad
  \partial^\nu \partial_\nu\, \phi_{\mu_1\dots\mu_s} \big|_\text{free e.o.m.} = 0\,.
\end{equation}
The relaxation of these conditions will allow to reconstruct the full off-shell counterpart of the TT vertices as in \cite{Manvelyan:2010jr,Francia:2016weg}.

\subsection{Vertex Generating Operators}
\label{sec:vertex-operators}

It is very convenient to contract the indices of the fields each with an auxiliary vector variable $a^\mu$,
\begin{align}
\phi^{(s)}(x,a)=\frac1{s!}\phi_{\mu_1\dots\mu_s}(x) a^{\mu_1}\cdots a^{\mu_s}\,.\label{phi}
\end{align}
This has several advantages: First, we do not have to tackle expressions with too many indices and secondly, the
tensor $\phi_{\mu_1 \cdots \mu_s}$ is by construction symmetric. We will also note later on that the complexity
of index contractions will be reduced a lot. For example, using the short-hand notation
$P^\mu = \partial_{x^\mu}$ and $A^\mu = \partial_{a^\mu}$, the relations in Eq.~(\ref{eq:phi-trace-div})
simplify to
\begin{equation}
  \label{eq:Fierz-phi}
  A^2\; \phi^{(s)} = 0\,, \qquad  A\cdot P \;\phi^{(s)} = 0\,, \qquad P^2\;
  \phi^{(s)}\big|_\text{free e.o.m.} = 0\,.  
\end{equation}
We call these relations collectively Fierz equations \cite{Fierz}.

Now, each $n$-point vertex $\mathcal{L}_n$ in Eq.~(\ref{eq:noether-exp}) is a product of $n$ massless bosonic
fields (and possibly derivatives thereof). But it has to be a Lorentz scalar, hence, all indices of the fields
(and of the derivatives) must be fully contracted. For now, let us concentrate on parity-even vertices -- we
consider parity-odd vertices in Section~\ref{sec:parity-odd-vertices}. 
Then, we can write $\mathcal{L}_n$ in the
following, very convenient way:
\begin{equation}
  \label{eq:L^n}
  \mathcal{L}_n (x) = \mathcal{V} \left( \prod_{i = 1}^n \phi_i(x_i,a_i) \right) \Bigg|_{\substack{x_i = x\\ a_i = 0}}\,.
\end{equation}
This needs some explanation:
\begin{itemize}
\item We use the notation set up in Eq.~(\ref{phi}) and drop the spin labels of the fields: $\phi_i$ is a
  spin $s_i$ field, $\phi_i = \phi^{(s_i)}$.
\item The term in brackets represents a function of the spacetime coordinates $x_i$ and the auxiliary vector
  variables $a_i$. The \textit{vertex generating operator} $\mathcal{V}$ performs the index contractions between
  the fields $\phi_i$ as follows: Let $P^\mu_i = \partial_{x^\mu_i}$ and $A^\mu_i = \partial_{a^\mu_i}$ as in
  Eq.~(\ref{eq:Fierz-phi}). Then, $\mathcal{V}$ must be a polynomial in the following commuting variables:
  \begin{align}
    \label{eq:zys}
    z_{ij} = A_i \cdot A_j \big|_{1 \leq i \leq j \leq n}, \qquad y_{ij} = A_i \cdot P_j \big|_{1 \leq i, j
    \leq n} , \qquad s_{ij} = P_i \cdot P_j \big|_{1 \leq i \leq j \leq n}\,.
  \end{align}
  The operator $z_{ij}$ induces a single contraction of indices between the fields $\phi_i$ and $\phi_j$,
  whereas $y_{ij}$ will take one index of the field $\phi_i$ and contract it with a derivative which acts on the
  field $\phi_j$. Finally, the operators $s_{ij}$ will introduce extra derivatives (a derivative of $\phi_i$ is
  contracted with a derivative of $\phi_j$); these are called Mandelstam variables.
\item Since all of the indices in $\mathcal{L}_n$ have to be contracted, we discard all terms that still contain
  at least one of the auxiliary variables when $\mathcal{V}$ acted on the terms in brackets. Thus, we set
  $a_i = 0$ in the end, which ensures that $\mathcal{L}_n$ is Lorentz invariant. Finally, we also set $x_i = x$.
  The splitting of the coordinates is useful to keep track of the derivatives acting on different fields, and
  has no physical consequences.
\end{itemize}

All in all, we translated the problem of `what is the most general form of the parity-even $n$-point vertex
$\mathcal{L}_n$' to the question `what is the most general form of the vertex generating operator $\mathcal{V}$
in the polynomial ring $\mathbb{R}[y_{ij}, z_{ij}|_{i \leq j}, s_{ij} |_{i \leq j}]$'. The connection between
Lagrangian $\mathcal{L}_n$ and operator $\mathcal{V}$ is given by Eq.~(\ref{eq:L^n}). We also ensured that
$\mathcal{L}_n$ is Lorentz invariant.

There are two questions arising now. First of all, there are equivalence relations for Lagrangians: e.g., two
Lagrangians that differ by a total derivative lead to the same action. We call them equivalent in this case.
What does this imply for the corresponding vertex generating operators? Secondly, how do we have to constrain
$\mathcal{V}$ such that $\mathcal{L}_n$ is gauge invariant? We present a general answer to these questions in
the remainder of this section and give more details in Sections~\ref{sec:case-2n-leq} and \ref{sec:case-n-}.

\subsection{Equivalence Relations for Vertex Generating Operators}
\label{sec:equiv-relat-vert}

We must take into account that different Lagrangians may describe the same theory. We say that they are
\textit{equivalent} in this case and evidently, we are only interested in $\mathcal{L}_n$ \textit{up to
  equivalence}. When we encode the Lagrangians via vertex generating operators, we need to introduce a notion of
equivalence for operators: \textit{vertex operators $\mathcal{V}$ and $\mathcal{V}'$ are
  equivalent, $\mathcal{V} \approx \mathcal{V}'$, iff the two Lagrangians $\mathcal{L}_n$
  and $\mathcal{L}_n'$, constructed from them via Eq.~(\ref{eq:L^n}) are also equivalent.} We are hence only interested
in $\mathcal{V}$ \textit{up to equivalence} and summarise the different kinds of equivalence relations in the
following.

The first kind of equivalence relations arises from field redefinitions $\phi_i \mapsto \phi_i + \delta \phi_i$,
where $\delta \phi_i$ is non-linear in the fields. These do not change the theory, but affect the
Lagrangian. For example, terms in $\mathcal{L}_2$ may contribute to $\mathcal{L}_n$ when the fields are
redefined non-linearly. But in this way, the $n$-point vertices only change by terms that vanish when the free
equations of motion are imposed. We say that two Lagrangians are equivalent, when they are related by such field
redefinitions and deduce from Eq.~(\ref{eq:L^n}) that we can choose $\mathcal{V}$ to be independent of $s_{ii}$.
Furthermore, we assume that $\mathcal{V}$ does not depend on $z_{ii}$ and $y_{ii}$ because the fields are
traceless and divergence-free.

Mathematically speaking, we impose the equivalence relations
\begin{equation}
  \label{eq:equiv1}
  y_{ii} \approx 0 , \qquad z_{ii} \approx 0 , \qquad s_{ii} \approx 0
\end{equation}
and deduce that each operator in the ideal
$\langle y_{ii}, z_{ii}, s_{ii} \rangle \subset \mathbb{R}[y_{ij}, z_{ij}|_{i \leq j}, s_{ij} |_{i \leq j}]$ is
equivalent to $0$. Hence, we can construct equivalence classes of vertex generating operators,
\begin{equation*}
  [\mathcal{V}] \in \frac{\mathbb{R}[y_{ij}, z_{ij}|_{i \leq j}, s_{ij}|_{i \leq j}]}{\langle y_{ii}, z_{ii},
    s_{ii}\rangle}\,. 
\end{equation*}
The quotient ring is isomorphic to the subring
$\mathcal{R} = \mathbb{R}\big[y_{ij}|_{i \neq j}, z_{ij}|_{i < j}, s_{ij}|_{i < j}\big]$,
\begin{equation*}
  \frac{\mathbb{R}[y_{ij}, z_{ij}|_{i \leq j}, s_{ij}|_{i \leq j}]}{\langle y_{ii}, z_{ii}, s_{ii}\rangle}
  \simeq 
  \mathcal{R} \subset \mathbb{R}[y_{ij}, 
  z_{ij}|_{i \leq j}, s_{ij}|_{i \leq j}]\,,
\end{equation*}
so we can choose the vertex generating operator as $\mathcal{V} \in \mathcal{R}$. In other words, we simply
dropped the dependence of $\mathcal{V}$ on $y_{ii}, z_{ii}$ and $s_{ii}$.

Secondly, acting with the operator $D^\mu = \sum_{j = 1}^n P_j^\mu$ on the term in brackets in
Eq.~(\ref{eq:L^n}) gives a total derivative in the Lagrangian. This does not change the action and hence, does
not affect the theory. Therefore, we impose the equivalence relations
\begin{equation}
  \label{eq:equiv2}
  A_i \cdot D = \sum_{j = 1}^n y_{ij} \approx 0,\qquad  P_i \cdot D = \sum_{j = 1}^n s_{ij} \approx 0 \,.
\end{equation}
These together generate an ideal $\mathcal{I}_D \subset \mathcal{R}$ and in the following, we consider equivalence
classes of vertex generating operators in the quotient ring
\begin{equation*}
  [\mathcal{V}] \in \frac{\mathcal{R}}{\mathcal{I}_D}\,.
\end{equation*}
As for the equivalence relations in Eq.~(\ref{eq:equiv1}), we could choose a convenient representative
$\mathcal{V}$ in $\mathcal{R}$, but it turns out to be better to keep the quotient ring structure for now.

A last equivalence stems from `Schouten identities', i.e.\ relations following from over-antisymmetrisation of
spacetime indices. These spacetime dimension-dependent identities are exact relations at the Lagrangian level. In
the polynomial ring $\mathcal{R}$, however, we forgot that we work in $d$ dimensions. Therefore, we have to impose Schouten
identities as equivalence relations for vertex generating operators,\footnote{Formally, let $\iota_d$ be the map
  \begin{align*}
    \iota_{d} : \qquad \qquad \mathcal{R} \qquad &\to \quad \mathbb{R}[P_i^{\mu},A_i^{\mu}] \\
    \mathcal{V}(z_{ij},y_{ij},s_{ij}) & \mapsto \mathcal{V}(A_i \cdot A_j, A_i \cdot P_j,P_i \cdot P_j)
  \end{align*}
  that replaces the operators $z_{ij}$, $y_{ij}$ and $s_{ij}$ by their definitions in Eq.~(\ref{eq:zys}).
  $\iota_d$ therefore reintroduces the operators $P_i$ and $A_i$ and hence, spacetime indices in $d$ dimensions in the vertex
  generating operator $\mathcal{V}$. The kernel $\iota_{d}^{-1} (0)$ of this map is what we call the ideal of Schouten identities in $d$ dimensions.} which form an
ideal $\mathcal{I}_S \subset \mathcal{R}$ as follows: Let $b = ( P_1 , \ldots , P_n, A_1, \ldots A_n )$ be a vector
of derivative operators and consider the symmetric $2n \times 2n $ matrix
\begin{equation}\label{defofB}
  \mathcal{B} = 
  \big(
  b_K \cdot b_L 
  \big)\big|_{K,L \in (1,\ldots, 2n)} = 
  \begin{pmatrix}
    \mathcal{S} & \mathcal{Y}^T \\ \mathcal{Y} & \mathcal{Z}
  \end{pmatrix}\,.
\end{equation}
Here, $\mathcal{S} = (s_{ij})$, $\mathcal{Y} = (y_{ij})$, $\mathcal{Z} = (z_{ij})$ are symmetric
$(n\times n)$-matrices with elements in $\mathcal{R}$. With the equivalence relations in Eq.~(\ref{eq:equiv1}), the
diagonal elements of $\mathcal{S}$, $\mathcal{Y}$ and $\mathcal{Z}$ vanish equivalently. We also keep in mind
that there are further equivalence relations from Eq.~(\ref{eq:equiv2}) which introduce a linear relation among the first $n$ rows (and columns) of $\mathcal{B}$, but we do not apply them right now.

Then, the ideal $\mathcal{I}_S$ is generated by all $(d + 1) \times (d+1)$ minors of $\mathcal{B}$. We show this in
a moment, but note first that this implies that $\mathcal{I}_S$ is trivial for $d \geq 2n - 1$. Indeed, in this
case, there is only one such minor, namely when equality holds. This minor is $\det \mathcal{B}$, which is equivalent to zero due to the equivalence relations in $\mathcal{I}_D$ (the first $n$ rows add up to a total derivative). Now we show that for $d < 2n - 1$, the above statement is true. Indeed, remove
$(2n - d-1)$ rows and columns from $\mathcal{B}$, such that only the rows
$K_1, \ldots, K_{d+1} \in (1,\ldots , 2n)$ and the columns $L_1, \ldots, L_{d+1} \in (1,\ldots , 2n)$ remain and
call the resulting $(d+1) \times (d+1)$-matrix $M$. Then,
\begin{align}
  \label{eq:schouten-proof}
  \det M &= \delta_{\nu_1}^{\mu_1} \cdots \delta_{\nu_{d+1}}^{\mu_{d+1}}\, B_{\mu_1 K_1} \cdots B_{\mu_{d+1} K_{d+1}}\,
           B^{\nu_1}_{[L_1} \cdots B^{\nu_{d+1}}_{L_{d+1}]} \nonumber \\
         &= \delta_{\nu_1 \cdots \nu_{d+1}}^{\mu_1 \cdots \mu_{d+1}}\, B_{\mu_1 K_1} \cdots B_{\mu_{d+1} K_{d+1}}\,
           B^{\nu_1}_{L_1} \cdots B^{\nu_{d+1}}_{L_{d+1}}
\end{align}
and acting with it on the term in brackets in Eq.~(\ref{eq:L^n}) yields a term in the Lagrangian with
over-antisymmetrised indices. On the other hand, each term in the Lagrangian with over-antisymmetrised indices
corresponds to a vertex generating operator $\mathcal{V}$ that contains a factor of the form on the rhs of
Eq.~(\ref{eq:schouten-proof}) for a certain set of indices $K_i,L_i \in (1,\ldots,2n)$. Hence,
$\mathcal{V} \in \mathcal{I}_S$.

At this step, it is convenient to introduce the notion of the \textit{level} of a Schouten identity. To this end, let us first define
the \textit{level} of the rows and columns of $\mathcal{B}$ as follows: The first $n$ rows and columns of
$\mathcal{B}$ are of level $0$ and all others are of level $1$. Furthermore, each $(d+1)\times(d+1)$-submatrix
$M$ of $\mathcal{B}$ that is obtained by removing rows and columns inherits those row and column levels from
$\mathcal{B}$. Then, the sum of row and column levels of $M$ equals the power of $A_i^\mu$ operators in
$\iota_d(\det M)$. This is what we call the \textit{level} of the Schouten identity $\det M \approx 0$. Denote by $I(k)$ the
ideal generated by all Schouten identities of level $k$, then we have
\begin{equation}
  \label{eq:grading}
  \mathcal{I}_S = \sum_{k = 0}^{2d + 2} I(k)\,,
\end{equation}
where again, $d$ denotes the spacetime dimension.

Now, we consider three cases:
\begin{itemize}
\item For large dimensions, $d \geq 2n-1$, as discussed before, there are no non-trivial Schouten identities at all (the only possible Schouten identities arise in the case $d=2n-1$, but they are zero up to total derivatives, so they are already contained in $\mathcal{I}_{D}$). This
  case is much simpler and we treat it separately in Section~\ref{sec:case-2n-leq}.
\item For large values of $n$, $d < n$, only the subideal $I(0)$ might be trivial (namely for $d + 1 = n$, where the level $0$ Schouten identities vanish up to a total derivative and thus are already contained in $\mathcal{I}_D$). Thanks to the variety of Schouten identities available, we
  are able to perform a lot of simplifications. We treat this case in Section~\ref{sec:case-n-}.
\item In the intermediate case $2n-2 \geq d \geq n$ only the ideals of level
  $2d - 2n + 4,\ldots, 2n$ are non-trivial. We will not study this case in full detail here, but a general characterisation of the corresponding vertices is given in Section~\ref{sec:discussion}.
\end{itemize}

All in all, we have now considered all possible equivalences for parity-even Lagrangians. Because of the freedom of field
redefinitions, we consider $\mathcal{V} \in \mathcal{R}$ and we divide out the ideals generated by total
derivatives ($\mathcal{I}_D$) and Schouten identities ($\mathcal{I}_S$),
  \begin{equation}
    \label{eq:[V]}
    [\mathcal{V}] \in \frac{\mathcal{R}}{\mathcal{I}} \,, \qquad\qquad \mathcal{I} = \mathcal{I}_S + \mathcal{I}_D\,.
  \end{equation}

\subsection{Imposing Gauge Invariance}
\label{sec:impos-gauge-invar}

Finally, we require that $\mathcal{L}$ is gauge invariant, i.e.\ that it satifies Eq.~(\ref{NE}). What does this imply
for the corresponding vertex generating operator $\mathcal{V}$? Note first that the second term in
Eq.~(\ref{NE}) vanishes when the free equations of motions are imposed. In other words, the requirement of gauge
invariance for the independent vertex structures reads
\begin{equation}
  \label{eq:gauge-0}
  \delta_k^{(0)} \mathcal{L}_n^{\,} \approx 0\,,
\end{equation}
where $\delta_k^{(0)}$ is the free gauge transformation of the field $\phi_k$ (see Eq.~(\ref{freegauge})).

The latter can be simplified by contracting the tensor for the gauge parameter in Eq.~(\ref{freegauge}) with
auxiliary vector variables $a^\mu$ as well,
\begin{equation}
  \epsilon^{(s-1)}(x, a)=\frac1{(s-1)!}\epsilon_{\mu_1\dots\mu_{s-1}}(x) a^{\mu_1}\cdots a^{\mu_{s-1}}\,.
\end{equation}
Again, we drop the spin index, $\epsilon_k = \epsilon^{(s_k - 1)}$, and the linearised gauge transformation of
the $k$-th field $\phi_k$ in Eq.~(\ref{freegauge}) reads
\begin{equation*}
  \delta^{(0)}_k \phi_k (x_k,a_k) = a_k\cdot P_k\;\epsilon_k (x_k,a_k),  \qquad \text{(no sum)}.
\end{equation*}
Note that this gauge transformation must be consistent with Eqs.~(\ref{eq:Fierz-phi}). We therefore impose the
Fierz equations also for the gauge parameter. 

All in all, we can now impose the restrictions for the vertex generating operators $\mathcal{V}$ from gauge
invariance, Eq.~(\ref{eq:gauge-0}):
\begin{equation*}
  \delta_k^{(0)} \mathcal{L}_n = \mathcal{V} \, a_k\cdot P_k \left( \epsilon_k(x_k,a_k) \prod_{1\leq i \leq n}^{i \neq k}
    \phi_i(x_i,a_i) \right) \Bigg|_{\substack{x_i = x\\ a_i = 0}}   \approx 0\,.
\end{equation*}
Since all the auxiliary vector variables $a_i$ are set to zero in the end, it immediately follows that
$\mathcal{L}_n$ is gauge invariant if and only if the corresponding vertex generating operator
$\mathcal{V} \in \mathcal{R}$ (via Eq.~(\ref{eq:L^n})) satisfies
\begin{equation}
  \label{eq:gauge-inv}
  \text{for all}\  k \in \{1,\ldots,n\} \,: \quad [\mathcal{V}, a_k \cdot P_k] =: D_k \mathcal{V} \in
  \mathcal{I}_S + \mathcal{I}_D\,. 
\end{equation}
Here, we defined the operators $D_k$ of gauge variations. These act as linear first-order differential operators
on the vertex $\mathcal{V}$:
\begin{align}
  D_k=\sum_{j=1}^n \Big( y_{jk}\frac{\partial}{\partial z_{kj}}+s_{kj}\frac{\partial}{\partial y_{kj}}\Big)\,.\label{gvo}
\end{align}

\section{The case $2n-1 \leq d$}
\label{sec:case-2n-leq}

We start with the case of sufficiently high spacetime dimensions where the classification of vertices is the
simplest because there are no Schouten identities and we only have to take into account total derivatives, hence, $\mathcal{I} = \mathcal{I}_D$.

\subsection{Gauge Invariants}
\label{sec:gauge-invariants}

To derive the $n$-th order independent vertices we first recall the constraints on the vertex generating operators
$y_{ij}\,, z_{ij}\,, s_{ij}$ in Eqs.~(\ref{eq:equiv1}) and (\ref{eq:equiv2}) and count the independent variables:
\begin{subequations}\label{constraints}
\begin{align}
y_{ii}\approx 0\,,\quad \sum_{j=1}^n y_{ij}\approx 0\,, \quad n(n-2)\; \text{variables}\; y_{ij}\,,\label{y}\\
z_{ij}=z_{ji}\,,\quad z_{ii}\approx 0\,,\quad \frac{n(n-1)}2\; \text{variables}\; z_{ij}\,,\\
s_{ij}=s_{ji}\,, \quad s_{ii}\approx 0\,,\quad  \sum_{j=1}^n s_{ij} \approx 0\,,\quad \frac{n(n-3)}{2}\; \text{variables}\; s_{ij}\,.
\end{align}
\end{subequations}
The vertex depends altogether on $2n(n-2)$ variables, and is subject to $n$ linear differential equations that
stem from Eqs.~(\ref{eq:gauge-inv}) and (\ref{gvo}).\footnote{Notice that the operators
  $D_{k}$ are consistent with these constraints~\eqref{constraints}, which means that $D_{k}$ acting on a
  constraint will lead to a constraint. Therefore we can leave the operators $D_k$ in the general form stated in
  Eq.~(\ref{gvo}) and do not need to express them in terms of a set of independent variables.} If these differential equations are linearly independent,
the solution should depend on $2n(n-2)-n=n(2n-5)$ variables.

For cubic vertices, $n=3$, this would give three invariants, while we know that the solution depends on four
invariants $y_{12}\,,\; y_{23}\,,\; y_{31}$ and $ G=y_{12}\,z_{23}+y_{23}\,z_{31}+y_{31}\,z_{12}\,$. The reason
is that the three equations are not linearly independent in that case:
$y_{12}\, D_1+y_{23}\, D_2+y_{31}\, D_3\approx 0$. Due to this relation, we have, e.g., the Yang-Mills cubic vertex
$V_3^{YM}=G$ and the Einstein-Hilbert cubic vertex $V_3^{EH}=G^2$.

On the other hand, one can easily see from Eq.~(\ref{gvo}) that the operators $D_k$ are linearly independent for
$n\geq 4$. Hence, the general form of the vertices should depend on $n(2n-5)$ invariants composed of
$s_{ij}\,,\; y_{ij}\,,\; z_{ij}\,$.

At this point, we introduce gauge invariant operators, which are more suitable as the building blocks of $n$-th
order vertices. These are given through the following variables:
\begin{align}
s_{ij}=s_{ji}\,\qquad & \frac{n(n-3)}2 \; \text{variables}\,,\\
c_{ij}=y_{ij}\,y_{ji}-s_{ij}\,z_{ij}=c_{ji}\,,\qquad & \frac{n(n-1)}2\; \text{variables}\,,\label{defc2} \\
c_{i,jk}=y_{ij}\,s_{ik}-y_{ik}\,s_{ij}=-c_{i,kj}\,, \qquad & \frac{n(n-2)(n-3)}2 \; \text{variables}\,.\label{defc3}
\end{align}
It is easy to show that these expressions are gauge invariant:
\begin{align}
D_k\, s_{ij}=0\,,\quad D_k\, c_{ij}=0\,,\quad D_k\, c_{i,jl}=0\,.
\end{align}

Counting the number of the variables $s_{ij}$ and $c_{ij}$ is straightforward. In order to count the number of
$c_{i,jk}$ variables, we count separately the number of choices for $i$ and the number of choices for the
antisymmetric pair $jk$ for a given $i$ and multiply them. Naively, we choose $i$ in $n$ possible ways, and the
antisymmetric pair $jk$ takes values in $\{i+1\,,\dots, i-2\; (\text{mod n})\}$, therefore takes
$\frac{(n-2)(n-3)}2$ values, hence the number of $c_{i,jk}$'s given above. These variables $c_{i,jk}$ are not
linearly independent though, satisfying the following relations:
\begin{align}
3\,c_{i,[jk}\, s_{i|l]}\equiv c_{i,jk}\, s_{il}+c_{i,kl}\, s_{ij}+c_{i,lj}\, s_{ik}=0\,\label{cijk redundancy}.
\end{align}
These naively are $\frac{n(n-2)(n-3)(n-4)}6$ many, given by multiplying the $n$ possible choices of $i$ and
$\frac{(n-2)(n-3)(n-4)}6$ choices of the antisymmetric triple $jkl$. But again, this counting is redundant, due
to linear relations between equations, involving different choices of $jkl$. These relations are also given by
adding another $s_{im}$ and antisymmetrising the four indices $jklm$. This chain of reducibility can be resummed
to get all linearly independent variables of $c_{i,jk}$. This is done by finding the number of possible values
of $jk$ antisymmetrised pairs that correspond to the independent variables, by summing up with changing signs
the numbers of components of antisymmetric tensors of $gl(n-2)$, starting from rank two\footnote{Remind, that the index $j$ takes $n-2$ independent values in $y_{ij}$ (and therefore in $c_{i,jk}$) due to \eqref{y}}:
\begin{align}
\sum_{i=2}^{n-2} (-1)^i {{n-2}\choose i}=n-3\,.
\end{align}

This means that the number of independent variables $c_{i,jk}$ is $n(n-3)$. We see that the variables $c_{i,jk}$
are redundant and we choose the following set of independent variables:
\begin{equation}\label{defYgenerald}
Y_{i}^{j} := c_{i,i+j\, i+1}\, ,
\end{equation}
where now $j=2,\dots, n-2$, taking $n-3$ possible values (indices are always meant modulo $n$). Thus, the number
of variables $Y_{i}^{j}$ is altogether $n(n-3)$. It is elementary to show that any other variable $c_{i,jk}$ can
be expressed through $Y_{i}^{j}$ using Eq.~\eqref{cijk redundancy}:
\begin{align}
\label{eq:Y(c)}
c_{i,jk}=\frac{c_{i,j\, i+1 }\,s_{ik}-c_{i,k\, i+1}\,s_{ij}}{s_{i\,i+1}}=\frac{Y_{i}^{j-i}\,s_{ik}-Y_{i}^{k-i}\,s_{ij}}{s_{i\,i+1}}\, .
\end{align}

Therefore altogether we have:
\begin{align}
\frac{n(n-3)}2+\frac{n(n-1)}2+n(n-3)=n(2n-5)\; \text{invariants}.
\end{align}
Given that the number of independent invariants $s_{ij}\,,\, c_{ij}\,,\, Y_{i}^{j}$ is the same as the number of
variables that should constitute the building blocks of $n$-th order independent vertices, it is already
tempting to conclude that the most general solution is an arbitrary function of these variables. We will show
this now, by allowing for dividing by Mandelstam variables and making the replacements
\begin{align}
z_{ij}=\frac1{s_{ij}}(y_{ij}\,y_{ji}-c_{ij})\,,
\end{align}
and, consecutively,
\begin{align}
y_{i i+j}=\frac1{s_{i i+1}} (y_{i i+1}\,s_{i i+j} - Y_{i}^{j})\,, \qquad \qquad j = 2,\ldots, n-2 \mod n\,,
\end{align}
expressing the vertex operator in terms of the variables $s_{ij},\, c_{ij},\, Y_{i}^{j}$ and $y_{i i+1}$. Correspondingly, the gauge variation in terms of these variables is generated by the operators
\begin{align}\label{Dksinglederivative}
D_k=s_{k k+1}\frac{\partial}{\partial y_{k k+1}}\,,
\end{align}
which turn into a single derivative. Therefore, the new gauge invariance equations for the vertex operator give:
\begin{align}
D_k \mathcal{V}(s_{ij},c_{ij},Y_{i}^{j},y_{i i+1})=s_{k k+1}\frac{\partial}{\partial y_{k k+1}}\mathcal{V}(s_{ij},c_{ij},Y_{i}^{j},y_{i i+1})\approx 0\,.\label{gie}
\end{align}
If we go to a set of independent variables, we can conclude that the $y_{ii+1}$-derivative is equal to zero, and
the vertex can be solely written in terms of the gauge invariant combinations $s_{ij},\, c_{ij},\, Y_{i}^{j}$. A
gauge invariant local vertex generating operator $\mathcal{V}$ in high enough dimension ($d \geq 2n - 1$) is
then in one-to-one correspondence to a polynomial in $s_{ij}$, $c_{ij}$, $Y_{i}^{j}$, allowing inverse powers of Mandelstam variables in such a way that $\mathcal{V}$ becomes polynomial in the variables $s_{ij}$,
$y_{ij}$ and $z_{ij}$, when re-expressing the combinations $c_{ij}$ and $Y_i^j$.

\subsection{Building blocks of vertices}

We have just shown that any gauge-invariant vertex $\mathcal{V}$ of order $n$ for $d\geq 2n-1$ can be rewritten
as a function of the invariants $c_{ij}$, $Y_{i}^{j}$ and $s_{ij}$. This function is polynomial in $c_{ij}$ and
$Y_i^j$, but can contain inverse powers of the Mandelstam variables $s_{ij}$.

In this subsection we address the question: `what is the most general form of this function if we assume that
the vertex is local?' First of all it is clear that any polynomial of $c_{ij}$, $Y_{i}^{j}$ and $s_{ij}$ defines
a local and gauge-invariant vertex. Now let us analyse the case that the vertex contains a single pole in one
$s_{ij}$ when written in terms of the invariants:
\begin{equation}
  \mathcal{V} = \frac{1}{s_{ij}} \mathcal{P} (c_{ij},Y_{i}^{j},s_{kl})\,.
\end{equation}
Here, we assume that the polynomial $\mathcal{P}$ does not explicitly depend on this specific $s_{ij}$. For
$\mathcal{V}$ to be local, the inverse of $s_{ij}$ has to be compensated by a term proportional to $s_{ij}$ that
arises when the invariants are rewritten in terms of $s_{ij},y_{ij},z_{ij}$. One can show that in this case,
$\mathcal{V}$ is a linear combination of
\begin{equation}
b_{ijk\ell}=\frac{1}{s_{ij}}\left(c_{ij}\,s_{ik}\,s_{j\ell} -c_{i,jk}c_{j,i\ell}\right) \quad \text{and}\quad \frac{1}{s_{ij}}\left(s_{ik}c_{i,j\ell}-s_{i\ell}c_{i,jk} \right)
\end{equation}
multiplied by polynomials in $c_{ij}$, $Y_i^j$ and the Mandelstam variables.\footnote{Note that $c_{i,jk}$ can
  be expressed as a polynomial in $Y$'s and Mandelstam variables via Eq.~(\ref{eq:Y(c)}).} The second
expression is simply equal to $c_{i,k\ell }$ (see~Eq.~\eqref{cijk redundancy}), so it is again a polynomial in
$s_{kl}$ and $c$ variables. The first one can be rewritten as
\begin{equation}
b_{ijk\ell} = \det \begin{pmatrix}
s_{ij} & s_{ik} & y_{ji}\\
s_{\ell j} & s_{\ell k}& y_{j\ell}\\
y_{ij} & y_{ik} & z_{ij}
\end{pmatrix} +s_{k\ell} c_{ij} \, .
\end{equation} 
Hence, up to a shift by a polynomial in Mandelstam and $c$ variables, the building block $b_{ijk\ell}$ can be
written as a determinant of a $3\times 3$-submatrix of the matrix $\mathcal{B}$ (see~Eq.~\eqref{defofB}). This
nicely fits with the observation that also the $c$ invariants are just minors of $\mathcal{B}$,
\begin{equation}
c_{ij} = - \det \begin{pmatrix}
s_{ij} & y_{ji}\\
y_{ij} & z_{ij}
\end{pmatrix} \quad ,\quad c_{i,jk} = \det \begin{pmatrix}
s_{ik} & s_{ij}\\
y_{ik} & y_{ij}
\end{pmatrix}\, .
\end{equation}
Notice that these minors as well as the $(3\times 3)$-example above have the property that each $(n+i)$-th row (column) of the second block  is accompanied by the corresponding ($i$-th) row (column) of the first block. This ensures gauge invariance because the $i$-th gauge variation transforms the $(n+i)$-th row (column) into the $i$-th row (column) leading to a vanishing determinant. Translating such a building block to the fields, the resulting expression is a pure curvature term: a tensor index of a field $i$ occurs in an antisymmetric combination with an index of a derivative acting on the field.

Of course all such minors can be written as polynomials in the $c$ invariants with negative powers of Mandelstam
variables allowed. This can be explicitly seen when in the determinant we add to the $(j+n)$-th column the
$j$-th column multiplied by $-\frac{y_{jj+1}}{s_{jj+1}}$, and similarly we add to the $(i+n)$-th row the $i$-th
row multiplied by $-\frac{y_{ii+1}}{s_{ii+1}}$. Then one arrives at
\begin{equation}
\det \begin{pmatrix}
(s_{ij}) & (y_{ji})\\
(y_{ij}) & (z_{ij})
\end{pmatrix}  = \det \begin{pmatrix}
(s_{ij}) & \Big(\frac{1}{s_{jj+1}} c_{j,ij+1} \Big)\\
\Big(\frac{1}{s_{ii+1}}c_{i,ji+1} \Big) & \Big(\frac{1}{s_{ij}s_{ii+1}s_{jj+1}} (c_{j,j+1\,i}c_{i,i+1\,j}-s_{ii+1}s_{jj+1}c_{ij}) \Big)
\end{pmatrix}\, .\label{InvariantMinor}
\end{equation}
Here, the labels $i$ and $j$ only run through the values that correspond to the rows and columns present in the minor that we are considering. 

There is one additional possibility due to the linear dependencies in $\mathcal{B}$: we can take the determinant
of the $(2n-1)\times (2n-1)$ submatrix that is obtained by deleting, e.g., the first row and column. This is
still gauge invariant because the gauge transformation with respect to the variables of the first field
transforms the first row of the second block into a linear combination of the $n-1$ rows of the first block, and
the determinant still vanishes. Expressed in terms of fields, such a building block corresponds to a term of the
form
\begin{equation}
\delta^{[\mu_{2}\dotsb \mu_{2n}]}_{\nu_{2}\dotsb \nu_{2n}}\phi^{(1)}_{\mu_{n+1}}{}^{\nu_{n+1}}\partial_{\mu_{2}}\partial^{\nu_{2}}\phi^{(2)}_{\mu_{n+2}}{}^{\nu_{n+2}} \dotsb \partial_{\mu_{n}}\partial^{\nu_{n}}\phi^{(n)}_{\mu_{2n}}{}^{\nu_{2n}}\, ,\label{Lovelock}
\end{equation}
which is gauge invariant up to total derivatives. This Lovelock-type vertex can be generalised in a way, where one computes the determinant of the minor of $\cal B$ containing $n-1$ rows and columns from the first block and arbitrary number $m$ of rows and columns from the second block, but these do not introduce new building blocks.\footnote{By adding total derivatives they can be transformed to an expression of the type \eqref{InvariantMinor} where the $n-1$ rows (columns) of the first block contain the $m$ rows (columns)  corresponding to those of the second block.}

Note that also the Mandelstam variables $s_{ij}$ are $1\times 1$-minors. It is tempting to speculate that all
gauge invariant local vertices $\mathcal{V}$ can be written as polynomials in the types of minors of
$\mathcal{B}$ mentioned above. If this speculation is correct, then for a spin configuration
$s_1\geq s_2\geq \dots \geq s_n$ $(n\geq 4)$ the lowest number of derivatives in a parity-even local vertex is
$s_1+s_2+\dots + s_{n}-2\lfloor \frac{s_n}{2}\rfloor$, and is achieved only for $\bar s_i = s_i-2\lfloor \frac{s_n}{2}\rfloor$ ($i=1,\dots,n$) satisfying polygon inequalities: $\bar s_1\leq \bar s_2+\dots+\bar s_n$.
In fact, taking into account the results of Section \ref{sec:parity-odd-vertices}, we can make a stronger statement for the special case $d=2n-1$: if the polygon inequalities between quantities $\tilde s_i = s_i-s_n$ ($i=1,\dots,n-1$) are satisfied ($\tilde s_1\leq \tilde s_2+\dots + \tilde s_{n-1}$), the lowest number of derivatives in a vertex is $N(s_i)=s_1+\dots +s_{n-1}$, where the corresponding vertex is parity-odd for odd $s_n$. When these polygon inequalities are not satisfied, the number of derivatives in the local vertex will be higher than $N(s_i)$.

\section{Lower dimension: dealing with Schouten identities}
\label{sec:lowerd}

In the previous section we have discussed the gauge-invariant vertices when we do not have to consider Schouten identities. When we go to lower dimensions, the ideal of relations is enlarged from $\mathcal{I}_{D}$ to $\mathcal{I}_{D}+\mathcal{I}_{S}$. Gauge-invariant vertex generating operators for large dimensions still define gauge-invariant operators in lower dimensions, but a priori, enlarging the ideal could have two effects: First, inequivalent vertices become equivalent, and second, new vertices arise that are gauge-invariant only up to the now larger set of equivalence relations. We will show in the following that the latter possibility does not lead to new equivalence classes of vertices for $n\geq 4$, but that for all gauge-invariant vertex generating operators there are equivalent operators\footnote{As long as we can divide by Mandelstam variables.} which are gauge-invariant already without the use of Schouten identities.
While this holds for quartic and all higher vertices in arbitrary dimensions, for cubic vertices ($n=3$) dimension-dependent vertices appear precisely in dimension $d=3$ (studied in \cite{Mkrtchyan:2017ixk,Kessel:2018ugi}).\footnote{We would like to make a side remark here on the cubic vertices of Fronsdal fields in $d=2$, discussed in the Appendix B of \cite{Kessel:2018ugi}. It can be shown, that the vertex $(s,s,0)$ is also trivial (due to the Schouten identity $y_i\,y_{i+1}\,z_{i-1}\approx 0$ in the notations of \cite{Kessel:2018ugi}) for $s\geq 2$. The only non-trivial parity-even vertices remaining are thus $\cV_{(1,1,0)}=y_1\,y_2$, $\cV_{(s,s,1)}=y_3\,z_3^s$ and the current coupling $\cV_{(s,0,0)}=y_1^s$ that has the same form as in arbitrary dimensions (the latter was forgotten in \cite{Kessel:2018ugi}). Note, that the on-shell triviality is not a reason to exclude the vertices as long as they cannot be removed via a {\it local} field redefinition, even if we assume that Fierz equations (which are stronger than free Fronsdal equations) can be removed by a field redefinition. This subtlety is discussed in detail in \cite{Francia:2016weg}.}

To show this, we start with a vertex generating operator $\mathcal{V}$ as a polynomial in $s_{ij},y_{ij},z_{ij}$ that in $d$ dimensions is gauge invariant,
\begin{equation}
D_{k}\mathcal{V} \in \mathcal{I}_{D}+\mathcal{I}_{S}\, .
\end{equation}
In $\mathcal{V}$ we now express the variables $z_{ij}$ and $y_{ij}$ in terms of $c_{ij}$, $Y_{i}^{j}$ and $y_{ii+1}$,
\begin{equation}
\mathcal{V} = \mathcal{P}_{\mathcal{V}} (c_{ij},Y_{i}^{j},y_{ii+1})\, ,
\end{equation}
where $\mathcal{P}_{\mathcal{V}}$ is a polynomial in the given variables. We suppressed the dependence on Mandelstam variables, which can also occur with negative powers.
In these variables, the gauge variation $D_{k}$ is written as a derivative with respect to $y_{kk+1}$ as in Eq.~\eqref{Dksinglederivative}, so we have
\begin{equation}
D_{k}\mathcal{V}= s_{kk+1}\frac{\partial}{\partial y_{kk+1}} \mathcal{P}_{\mathcal{V}} (c_{ij}, Y_{i}^{j},y_{ii+1}) \in \mathcal{I}_{D}+\mathcal{I}_{S}\, .
\end{equation}
When we expand $\mathcal{P}_\mathcal{V}$ in powers of $y_{12}$,
\begin{equation}
\mathcal{P}_{\mathcal{V}} (c_{ij}, Y_{i}^{j},y_{ii+1})  = \sum_{k=0}^{K} q_{k} (c_{ij}, Y_{i}^{j},y_{23},\dotsc ,y_{n1}) (y_{12})^{k} \, ,
\end{equation}
we apply $(D_{1})^{K}$ to the expression and obtain 
\begin{equation}
K! (s_{12})^{K} q_{K} \in \mathcal{I}_{D}+\mathcal{I}_{S}\, .
\end{equation} 
When we allow ourselves to divide by Mandelstam variables, we conclude that 
\begin{equation}
q_{K} \in \frac{1}{(s_{12})^{K}} \left(\mathcal{I}_{D}+\mathcal{I}_{S} \right)\, .
\end{equation}
Similar relations can be found for all other terms in the expansion in $y_{12}$ and also in the other variables $y_{ii+1}$. Hence, we find that
\begin{equation}
\mathcal{V} - \mathcal{P}_{\mathcal{V}}(c_{ij}, Y_{i}^{j},y_{ii+1})\Big|_{y_{ii+1}=0} \in \frac{1}{\Delta} \left(\mathcal{I}_{D}+\mathcal{I}_{S} \right)\, ,
\end{equation}
where $\Delta$ is a product of powers of Mandelstam variables. Therefore, $\mathcal{V}$ is equivalent to an operator depending only on $c_{ij}$ and $Y_{i}^{j}$ which already defines a gauge invariant vertex operator without the need of Schouten identities.

We conclude that in all dimensions, vertex generating operators can be expressed in terms of the operators identified for large dimensions. The main task for lower dimensions is therefore to work out explicitly the equivalences between such operators that are induced by Schouten identities. Here, the case of low dimensions, $d<n$, is special because many Schouten identities arise that reduce the independent equivalence classes considerably. This will be discussed in detail in the subsequent section. The identifications in the intermediate case will be stated in the discussion in Section~\ref{sec:discussion}.
\smallskip

In the remainder of this section we give a heuristic geometric argument why generically one does not expect new vertices to appear when we lower the dimension. 
In the sense of algebraic geometry, the ideal $\mathcal{I} = \mathcal{I}_S + \mathcal{I}_D$ defines a variety
$V (\mathcal{I})$ as the zero-set of the polynomials contained in $\mathcal{I}$. If $\mathcal{I}$ was a prime
ideal, we could think of the ring $\mathcal{R}/\mathcal{I}$ as the ring of polynomial functions on this variety.
The gauge variations $D_{k}$ define $n$ vector fields on this variety, and we are looking for functions on
$V(\mathcal{I})$ that are constant along the vector fields. When we enlarge the ideal to
$\mathcal{I}'\supset \mathcal{I}$ by going from higher to lower dimensions where new Schouten identities occur,
we concentrate on a subvariety $V (\mathcal{I}')$ of $V (\mathcal{I})$. Generically, if the vector fields do not
degenerate on this subvariety, functions that are constant along $D_{k}$ on $V (\mathcal{I}')$ can be lifted to
constant functions on $V (\mathcal{I})$.

The above argument only gives a very rough picture, because apart from the possible degeneration of the vector
fields, there are two subtleties: First, as it was said, the argument only applies to prime ideals, but the
ideals that occur are usually not prime; secondly, there could be constant polynomials on $V (\mathcal{I}')$ whose lifts
to $V(\mathcal{I})$ are not polynomial. Therefore, this picture can only be seen as a heuristic explanation why generically we do not
expect new gauge invariant vertices to appear when we lower the dimension.

\section{The case $n > d$}
\label{sec:case-n-}

In this chapter, we find general restrictions for gauge invariant $n$-point vertices with $n > d$. Our result is
a simple characterisation of equivalence classes $[\mathcal{V}] \in \mathcal{R} \slash \mathcal{I}$ for vertex
generating operators. The results are summarised in Section~\ref{sec:restrictions}.

As discussed in Section~\ref{sec:equiv-relat-vert}, we have the full set of Schouten identities at hand in
order to find a simple representative $\mathcal{V}$ for a given vertex. This has the advantage that a lot of
simplifications are possible.
On the other hand, the structure of the set of Schouten identities is complicated, and the number of linearly independent Schouten identities in the polynomial ring, $\frac{1}{2}\binom{2n-1}{d+1}\left(\binom{2n-1}{d+1}+1 \right)$ for $n\geq 4$, is large and rapidly growing with $n$.
This problem was solved in \cite{Fredenhagen:2019hvb} for $d = 3$ by observing that many Schouten identities
become dependent when multiplied with an appropriate product $\Delta$ of Mandelstam variables. By multiplying a
given vertex $\mathcal{V}$ with $\Delta$, the remaining independent Schouten identities can be used to deduce
strong constraints for the vertex $\mathcal{V}$ itself. Essentially, one can treat the Mandelstam variables in
the manipulations like numbers and also divide by them. This concept can be also employed in higher dimensions.

Formally, to be able to divide by certain combinations of Mandelstam variables, we introduce the ring of
fractions, $M^{-1}\mathcal{R}$. Here, $M$ is a multiplicatively closed set containing all (finite) products of
\textit{non-zero minors of the submatrix $\mathcal{S}$ of $\mathcal{B}$} (see Eq.~\eqref{defofB}): these are the
expressions we want to divide by. More explicitly, let $\Mi(\mathcal{S})$ be the set of non-zero minors of
$\mathcal{S}$~\footnote{First, non-zero minors of order one are just the Mandelstam variables $s_{ij}$ with
  $i \neq j$. Secondly, all minors of order $2,3,\dots, d$ are generically non-zero -- even when the
  equivalence relations in Eq.~(\ref{eq:equiv2}) are applied. Finally, all minors of order greater than $d$ do
  vanish due to Schouten identities. Hence, $\Mi(\mathcal{S})$ consists of all
  $(2\times2)\,,\; (3\times3)\,,\; \dots (d \times d)$ subdeterminants of $\mathcal{S}$ as well as the
  Mandelstam variables $s_{ij}$ with $i \neq j$.} and let $M=\Mon [\Mi(\mathcal{S})]$ be the set of monomials in
these minors. Then, the ring of fractions consists of formal quotients,
\begin{equation}
M^{-1}\mathcal{R} = \Big\{ \frac{r}{\Delta} \, \Big|\, \Delta \in M,\ r\in \mathcal{R} \Big\}\, ,
\end{equation}
with the obvious rules for addition and multiplication. As also $1\in M$, we can identify $\mathcal{R}$ via $r \mapsto \frac{r}{1}$ as subring of $M^{-1}\mathcal{R}$. The ideal $\mathcal{I}=\mathcal{I}_{S}+\mathcal{I}_{D}\subset \mathcal{R}$ can then be seen as a subset of $M^{-1}\mathcal{R}$ which generates an ideal $\mathcal{I}_{M}$ in $M^{-1}\mathcal{R}$. Using the embedding of $\mathcal{R}$ into $M^{-1}\mathcal{R}$, we have an induced map of the quotient rings,
\begin{equation}
i_{M}:\frac{\mathcal{R}}{\mathcal{I}} \to \frac{M^{-1}\mathcal{R}}{\mathcal{I}_{M}}\, .
\end{equation}

As we will argue below, this map is injective, and therefore we can characterise equivalence classes of vertices uniquely by equivalence classes in the ring of fractions. The crucial observation is now that in $M^{-1}\mathcal{R}$ many of the generators of the ideal become dependent, so that $\mathcal{I}_{M}$ has a simple set of generators.

This section is structured as follows. In Section~\ref{sec:ideal} we find a simple set of generators for
$\mathcal{I}_{M}$, which enables us to find a convenient representative of $[\mathcal{V}]$ in the quotient of
the ring of fractions in Section~\ref{sec:using-equiv-relat}. We then impose gauge invariance in
Section~\ref{sec:gener-restr-from}, which leads to strong restrictions on the vertex $\mathcal{V}$. In $d=3$,
these restrictions completely rule out independent vertices (as reported in \cite{Fredenhagen:2019hvb}), in
higher dimensions the restrictions are less strict, and we discuss them in Section~\ref{sec:restrictions}. In
order to make the structure of this paper better accessible, we collect some proofs in Section~\ref{sec:proofs}.

Before we proceed, we want to show that $i_{M}$ is indeed injective. If $i_{M} ([\mathcal{V}]) =[0]$, this means
that $\mathcal{V}\in \mathcal{R}\cap \mathcal{I}_{M}$. Then, there is some $\Delta \in M$ such that
$\Delta \mathcal{V}\in \mathcal{I}$. If $\Delta \mathcal{V}$ defines a trivial vertex, then also
$\mathcal{V}$ corresponds to a trivial vertex, which can be seen in Fourier space, where the operators $s_{ij}$
are numbers. In particular, the polynomial $\Delta$ is non-zero on the subvariety defined by $k_{i}^{2}=0$ and
$\sum k_{i}=0$. Now, if $\Delta \mathcal{V}$ defines a trivial vertex, then
\begin{equation}
  \Delta \mathcal{V} \prod_i \widehat{\phi}_i(k_i,a_i) \Big|_{a_i = 0}
\end{equation}
vanishes on this subvariety. The factor $\Delta$ is non-vanishing almost everywhere. Hence, since $\mathcal{V}$ only
depends polynomially on $k_{i}^{\mu}$, $\mathcal{V}$ applied on the fields $\widehat{\phi}_{i}$ must vanish. So
we conclude that $\mathcal{V}\approx 0$, hence $[\mathcal{V}] = [0]$.

\subsection{A Minimal Generating Set of Schouten Identities}
\label{sec:ideal}

In this section, we find a simple set of generators for the ideal $\mathcal{I}_M$ in two steps. First, any
Schouten identity multiplied with a certain $\Delta \in M=\Mon [\Mi(\mathcal{S})]$ is an element in the ideal
generated by the equivalence relations in Eq.~(\ref{eq:equiv2}) and all Schouten identities up to
level~2~\footnote{This proof relies on the fact that $n > d$.} (recall the notion of level introduced in the
paragraph before Eq.~(\ref{eq:grading})). In other words,
\begin{equation}
  \label{eq:MAIN} \text{there exists}\  \Delta \in \Mon [\Mi(\mathcal{S})] \ \text{such that} \ \Delta \cdot \mathcal{I}_S \subset
  \sum_{k = 0}^2 I(k) + \mathcal{I}_D \,.
\end{equation}
We show this in Section~\ref{sec:scho-ident-revis}. This observation implies that in the ring of fractions where
we are allowed to divide by $\Delta $, we need far less generators for the Schouten identities.

In order to perform the second step, we introduce some more notations: First,
\begin{equation}
  \label{eq:N_ij}
  N_{ij} =  
  \begin{pmatrix}
    s_{ij} & \cdots &  s_{i j+d-1} \\
    \vdots & \ddots &  \vdots \\
    s_{i+d-1 \, j}&\cdots & s_{i+d-1 j+d-1}
  \end{pmatrix}
\end{equation}
is a $d\times d$ submatrix of $\mathcal{S}$, hence, $\det N_{ij} \in \Mi(\mathcal{S})$ and $N_{ij}$ has full
rank. Secondly, let $B_1(i,j)$ with $i,j = 1,\ldots,n$ be the following $(d+1)\times(d+1)$ submatrix of
$\mathcal{B}$: It contains the rows and columns $i,i+1,\ldots,i + d - 1$ (modulo $n$) as well as another row $j$
and the column $i + n$. Hence,
\begin{equation}
  \label{eq:B1}
  \det B_1(i,j) = \det
  \left(
    \begin{array}{cccc}
      &&&0 \\
      &&& y_{ii+1}\\
      \multicolumn{3}{c}{\smash{\raisebox{.5\normalbaselineskip}{$N_{ii}$}}} &\vdots \\
      &&& y_{ii+d-1}\\
      s_{ji} & \cdots & s_{j i+d-1} & y_{ij}
    \end{array}
  \right)
  \in I(1)\,.\footnote{Note that this is true for all $j = 1,\ldots, n$. If for example $j = i$, then
    $\det B_1(i,j) = 0 \in I(1)$.}
\end{equation}
Finally, let $B_2(i,j)$ with $i,j = 1,\ldots,n$ be the $(d+1)\times(d+1)$ submatrix of $\mathcal{B}$ containing
the rows $i,i+1,\ldots,i + d - 1$ (modulo $n$) and $i+n$, as well as the columns $j,j+1,\ldots,j + d - 1$
(modulo $n$) and $j+n$. Hence,
\begin{equation*}
  \det B_2(i,j) = \det
  \left(
    \begin{array}{cccc}
      &&& y_{ji}\\
      \multicolumn{3}{c}{\smash{\raisebox{.0\normalbaselineskip}{$N_{ij}$}}} &\vdots \\ 
      &&& y_{j i+d-1}\\
      y_{ij} & \cdots & y_{i j+d-1} & z_{ij}
    \end{array}
  \right)
  \in I(2)\,.
\end{equation*}
With these notations, we show in Section~\ref{sec:schouten-ideal-2} that there exists $\Delta \in M=\Mon [\Mi(\mathcal{S})]$ such that
\begin{equation}
  \label{eq:I-Delta}
   \Delta \cdot (\mathcal{I}_S +
  \mathcal{I}_D) \subset I(0) + \left\langle \sum_{k = 1}^n s_{ik}\,,\, \det B_1(i,j)\,,\, \det B_2(i,j) \; \Big| \; i,j
    = 1,\ldots, n \right\rangle\,.
\end{equation}
Denote the family of generators of $I (0)$ by $(\det B_0(A))$, where $A$ labels the different equivalence
relations. Then, we can conclude that $\mathcal{I}_{M}$ is generated as
\begin{equation}
\mathcal{I}_{M} = \left\langle \sum_{k = 1}^n s_{ik}\,,\,(\det B_0 (A))\, ,\, \det B_1(i,j)\,,\, \det B_2(i,j) \; \Big| \; i,j
    = 1,\ldots, n \right\rangle\,.
\end{equation}

\subsection{The choice of Representative}
\label{sec:using-equiv-relat}

Now, let us investigate the relevant ideal $\mathcal{I}_M$ in order to choose a convenient representative
for $\mathcal{V}$ in its equivalence class
$[\mathcal{V}] \in M^{-1}\mathcal{R} \slash \mathcal{I}_M$.

We start by considering the Schouten identities $\det B_2(i,j) \in I(2)$, with $i \neq j$. Using a Laplace
expansion along the last column, they read
\begin{equation}
  \label{eq:simplify-z}
  0 \approx \det B_2(i,j) = z_{ij} \det N_{ij} + \text{terms that do not contain any } z_{kl}\,.
\end{equation}
Since $\det N_{ij} \in \Mi(\mathcal{S})$, we can divide by it in $M^{-1}\mathcal{R}$, and express $z_{ij}$ by an expression independent of any $z_{kl}$. Hence, we may choose the representative of $[\mathcal{V}]$ to be independent of $z_{ij}$. In the same way,
the Schouten identities $\det B_1(i,j) \in I(1)$ take the form
\begin{equation*}
  0 \approx \det B_1(i,j) = y_{ij} \det N_{ii} + p(s_{ij},y_{i i+1} ,\ldots, y_{i i+d-1}) \,.
\end{equation*}
Here, the polynomial $p$ only depends on $y_{i i+1} ,\ldots, y_{i i+d-1}$~\footnote{The indices are considered
  modulo $n$.} and the Mandelstam variables. Using these Schouten identities, we can replace all of the
operators $y_{ij}$ in $\mathcal{V}$ except for $y_{ii+1},\ldots,y_{ii+d-1}$.
  
Finally, we perform a change of variables in $\mathcal{V}$. Similarly to Eq.~\eqref{defYgenerald} we introduce the combinations
\begin{equation}
  \label{eq:Y}
  Y_i^j = s_{i i+1} y_{i i+j} - s_{i i+j} y_{i i+1} \qquad \text{for } j = 2,\ldots , d - 1\,,
\end{equation}
and replace all $y_{ii+2},\dotsc ,y_{ii+d-1}$ in terms of these variables and $y_{ii+1}$. This can be done,
because $s_{ii+1} \in M$ and we can divide by it in $M^{-1} \mathcal{R}$. We arrive at
\begin{equation}
  \label{eq:Q}
  \mathcal{V} \approx \mathcal{P}_\mathcal{V}(y_{ii+1}, Y_i^j,s_{ij})\,, 
\end{equation}
where $\mathcal{P}_\mathcal{V}$ is a polynomial in $y_{ii+1}$, $Y_i^j$ and the Mandelstam variables (with coefficients
that can contain inverse powers of elements in $\Mi (\mathcal{S})$). More explicitly, we can see $[\mathcal{V}]$ as an
element in the quotient
\begin{equation}
\label{eq:V-in-MR}
  [\mathcal{V}] \in \frac{M^{-1}\mathbb{R}\left[y_{ii+1}, Y_i^j , s_{ij}\right]}{\left\langle (\det B_0 (A))\,,\,\sum_{j =
          1}^n s_{ij}\,,\,  \det B_2(i,i)\;\big|\; i = 1,\ldots, n \right\rangle}\,.
\end{equation}
There are several reasons to introduce the $Y_i^j$ variables. First, they are the gauge invariant combinations
of the $y_{ij}$ variables -- we have discussed this already in Section~\ref{sec:case-2n-leq} and it will become important in Section~\ref{sec:gener-restr-from}. Secondly, the remaining level-2
Schouten identities $\det B_2(i,i)$ can be written solely in terms of the $Y_i^j$'s and the Mandelstam
variables, and they do not depend explicitly on $y_{ii+1}$. We show this in the remainder of this section: For this purpose, consider
\begin{equation*}
  s_{ii+1}^2 \det B_2(i,i) = \det
  \left(
    \begin{array}{ccccc}
      &&&&0 \\
      &&&& s_{ii+1} y_{ii+1}\\
      \multicolumn{4}{c}{\smash{\raisebox{.5\normalbaselineskip}{$N_{ii}$}}} &\vdots \\
      &&&& s_{ii+1}y_{ii+d-1}\\
      0 & s_{ii+1} y_{ii+1} & \cdots & s_{ii+1} y_{i i+d-1} & 0
    \end{array}
  \right)\,.
\end{equation*}
The determinant of the matrix does not change when $y_{ii+1}$ times the first row is subtracted from the last
one and $y_{ii+1}$ times the first column is subtracted from the last one. Hence, using the definition of
$Y_i^j$ in Eq.~(\ref{eq:Y}), we find
\begin{equation*}
  s_{ii+1}^2 \det B_2(i,i) = \det
  \left(
    \begin{array}{cccccc}
      &&&&&0 \\
      &&&&& 0\\
      \multicolumn{5}{c}{\smash{\raisebox{.0\normalbaselineskip}{$N_{ii}$}}} &Y_i^2 \\
      &&&&& \vdots\\
      &&&&& Y_i^{d-1}\\
      0 & 0 & Y_i^2 & \cdots & Y_i^{d-1} & 0
    \end{array}
  \right)
  = - \sum_{j,k = 2}^{d-1} Y_i^j \left( \text{adj}\,N_{ii} \right)_{jk} Y_i^k =: q_2^i(Y_i^j,s_{jk})\,.
\end{equation*}
Here, we used a Laplace expansion along the last row and column, and $ \text{adj}\,N_{ii}$ denotes the adjugate matrix of $N_{ii}$. The resulting polynomials $q_2^i$ are quadratic
in the $Y_i^j$ variables with coefficients that still depend on the Mandelstam variables. However, the $q_2^i$'s
are independent of $y_{ii+1}$. We comment on their structure in Section~\ref{sec:restrictions}. All in all, we
can replace the generators $\det B_2(i,i)$ by $q_2^i$ because we are allowed to divide by Mandelstam variables.
Hence, we have the following result:
\begin{equation*}
  [\mathcal{V}] \in \frac{M^{-1}\mathbb{R}\left[y_{ii+1}, Y_i^j , s_{ij}\right]}{\left\langle (\det B_0 (A))\,,\,\sum_{j =
        1}^n s_{ij}\,,\,  q_2^i\;\big|\; i = 1,\ldots, n \right\rangle}\,.
\end{equation*}

\subsection{General Restrictions from Gauge Invariance}
\label{sec:gener-restr-from}

With the results of the previous sections, we now show that the polynomial $\mathcal{P}_\mathcal{V}$ introduced in
Eq.~(\ref{eq:Q}) can be chosen to be independent of $y_{i i+1}$ if the operator $\mathcal{V}$ corresponds to a gauge
invariant Lagrangian $\mathcal{L}_n$. From now on, we will always consider $\mathcal{V}$ as an element in the bigger ring of fractions.

Starting from Eq.~(\ref{eq:gauge-inv}) and using that the operators $a_k\cdot P_k$ commute with all Mandelstam
variables, we find that a gauge invariant vertex $\mathcal{L}_n$ requires
\begin{equation*}
  \text{for all}\  k \in \{1, \ldots n\} \,: \quad [\mathcal{V} , a_k \cdot P_k] \in \mathcal{I}_M\,, 
\end{equation*}
where $\mathcal{L}_n$ and $\mathcal{V}$ are related via Eq.~(\ref{eq:L^n}). Now, since the ideal $\mathcal{I}_M$
is gauge invariant, $[\mathcal{I}_M, a_k \cdot P_k]\subset \mathcal{I}_{M}$, we deduce that the polynomial in
Eq.~(\ref{eq:Q}) satisfies
\begin{equation*}
  [\mathcal{P}_\mathcal{V}, a_k \cdot P_k ] \in \left\langle (\det B_0 (A))\,,\,\sum_{j =
      1}^n s_{ij}\,,\,  q_2^i(Y_i^j)\;\big|\; i = 1,\ldots, n \right\rangle\,.
\end{equation*}
With
\begin{equation*}
  [y_{ii+1}, a_k \cdot P_k] = \delta_{ik} s_{i i +1}  \qquad \Rightarrow \qquad [Y_i^j , a_k \cdot P_k] = \delta_{ik}
  \left(s_{ii+j} s_{i i + 1} - s_{ii+1} s_{i i+j} \right) = 0\,,
\end{equation*}
it follows immediately that 
\begin{equation*}
  \text{for all}\  k = 1,\ldots, n \,:\quad s_{kk+1} \partial_{y_{kk+1}} \mathcal{P}_\mathcal{V} \in \left\langle (\det B_0 (A))\,,\,\sum_{j =
      1}^n s_{ij}\,,\,  q_2^i(Y_i^j)\;\big|\; i = 1,\ldots, n \right\rangle\,.
\end{equation*}
The generators of the ideal on the rhs do not depend on $y_{ii+1}$. We conclude that $\mathcal{P}_\mathcal{V}$ can be chosen to be 
independent of $y_{ii+1}$. More explicitly,
\begin{equation}
  \label{eq:result-schoutens}
  \mathcal{V} \approx \mathcal{P}_\mathcal{V} (Y_i^j\,,\,s_{ij})\,, \qquad[\mathcal{V}] \in \frac{M^{-1}\mathbb{R}\left[Y_i^k , s_{ij}\right]}{\left\langle (\det B_0 (A))\,,\,\sum_{j = 1}^n s_{ij} \,,\, q_2^i(Y_i^j) \;|\; i = 1,\ldots , n \right\rangle}\,.
\end{equation}

\subsection{Restrictions for $\mathcal{V}$}
\label{sec:restrictions}

Let us summarise our results. Eq.~(\ref{eq:result-schoutens}) states that each gauge invariant vertex
$\mathcal{V}$ is equivalent to a vertex $\mathcal{P}_\mathcal{V}$ that does only depend on Mandelstam variables and
$Y_i^j$. In particular, translating back to the vertex in terms of $P_i^\mu$ and $A_i^\mu$ operators, we have
the following relation:
\begin{equation*}
  \iota_d (Y_i^j) = 2 P_i{}_\mu A_i{}_\nu 
  P_{[i+1}^{\mu} P_{i+j]}^{\nu} = 2 P_i{}_{[\mu} A_{|i|}{}_{\nu]} 
  P_{i+1}^\mu P_{i+j}^\nu\,.
\end{equation*}
Now, in the vertex generated by $\mathcal{P}_\mathcal{V}$, an index of the $i$th field is only generated by $A_i^\mu$ via
a corresponding $Y_i^j$. Hence, the $i$th field enters the Lagrangian via a curvature term (each index of the
field is antisymmetrised with an index of a partial derivative acting on it). We deduce that $\mathcal{P}_V$ generates a
Lagrangian that can be written solely in terms of curvature terms.

The drawback of this analysis is that we do not control locality on the way to this result. $\mathcal{P}_\mathcal{V}$ might not
have a local form, since it can have inverse powers of Mandelstam variables. We can only say that for each
gauge invariant vertex (generated by $\mathcal{V}$), there is a $\Delta \in M$ such that $\Delta \mathcal{V}$
can be written only in terms of curvatures.

Much stricter conditions can be found in three dimensions \cite{Fredenhagen:2019hvb}. In that case, there is
only one $Y_i^j$ and the corresponding Schouten identity is $q_2^i = - s_{ii+1}^2 (Y_i^2)^2$. Hence,
$\det B_2(i,i) = (Y_i^2)^2 \approx 0$ and $\mathcal{P}_\mathcal{V}$ is only linear in $Y_i$. One can then deduce that
$\mathcal{V}$ itself is at most linear in each of the operators $A_i^\mu$, which means that the corresponding
vertex $\mathcal{L}_n$ contains no higher-spin fields at all. Indeed, in $d = 3$ there are simply no on-shell non-trivial curvature terms for higher-spin fields.

Our analysis also applies to the case $d=2$. Here, no $Y_{i}^{j}$ remain, and therefore there are no independent vertices for $n\geq 4$ involving massless Fronsdal fields of spin $s\geq 1$.

\subsection{Proofs}
\label{sec:proofs}

\subsubsection{Proof of Eq.~(\ref{eq:MAIN})}
\label{sec:scho-ident-revis}

Let $\det M = 0$ be a Schouten identity that stems from a $(d+1)\times(d+1)$-submatrix $M$ of $\mathcal{B}$ such
that $\det M \not{\in} \mathcal{I}_D$. Let $r$ ($s$) be the number of level-0 rows (columns) of $M$.
Furthermore, let $\bar r$ ($\bar s$) be the number of level-1 rows (columns) of $M$. Hence,
$r + \bar r = s + \bar s = d+1$. Without loss of generality, we assume $r \geq s$.\footnote{If $r < s$, we
  choose $M^T$ instead of $M$, which yields the same Schouten identity $\det M^T = \det M$. $M^T$ is a submatrix
  of $\mathcal{B}$ as well because $\mathcal{B}$ is symmetric.} Furthermore, let $\bar s \geq 2$, hence, the
level of the Schouten identity $\det M = 0$ is $\bar r + \bar s \geq 2$. In particular, equality holds if and
only if $\bar s = 2$ and $\bar r = 0$.

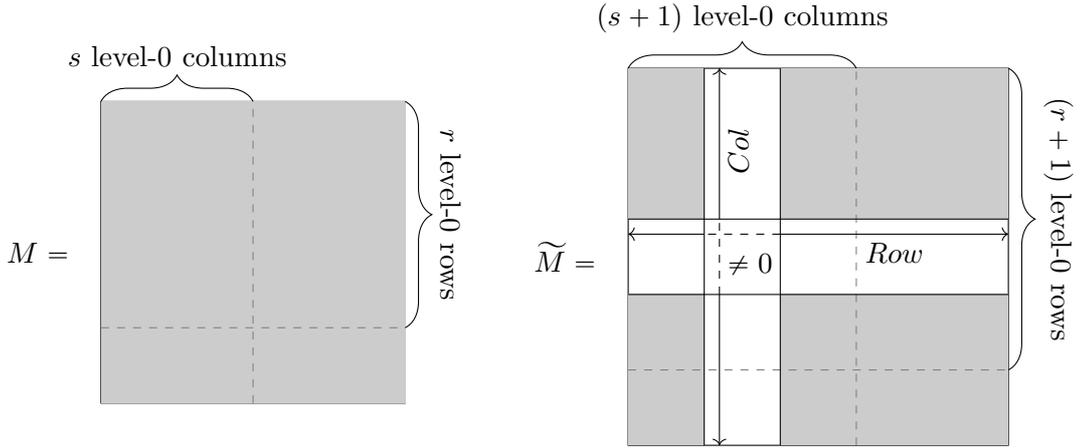
\begin{figure}[t]
  \centering
\begin{minipage}[!]{0.45\linewidth}
  
\begin{tikzpicture}
\matrix[draw,inner sep=0pt] (S) [matrix of math nodes,nodes={outer sep=0pt,minimum width=10mm,minimum
  height=10mm}]
{|[fill=black!20]|  & |[fill=black!20]| & |[fill=black!20]| & |[fill=black!20]|\\
|[fill=black!20]|  & |[fill=black!20]| & |[fill=black!20]| & |[fill=black!20]|\\
|[fill=black!20]|  & |[fill=black!20]| & |[fill=black!20]| & |[fill=black!20]|\\
|[fill=black!20]| & |[fill=black!20]| & |[fill=black!20]| & |[fill=black!20]|\\
};

\draw[dashed,gray] (S-4-2.south east) -- (S-1-2.north east);
\draw[dashed,gray] (S-4-1.north west) -- (S-4-4.north east);

\draw[decorate,decoration={brace,amplitude=10pt}] (S-1-1.north west) -- (S-1-3.north west);
\draw[decorate,decoration={brace,amplitude=10pt},rotate=90] (S-1-4.north east) -- (S-4-4.north east);
\node[above,yshift=2ex,xshift=-3ex] (col) at (S-1-2.north) {$s$ level-0 columns};
\node[rotate=270,above,yshift=2ex] (row) at (S-2-4.east) {$r$ level-0 rows};

\node[xshift=-5ex] (M=) at (S-3-1.north west) {$M=$};
\end{tikzpicture}

\end{minipage}
\begin{minipage}[!]{0.45\linewidth}

\begin{tikzpicture}
\matrix[draw,inner sep=0pt] (S) [matrix of math nodes,nodes={outer sep=0pt,minimum width=10mm,minimum
  height=10mm}]
{|[fill=black!20]| & \ & |[fill=black!20]| & |[fill=black!20]| & |[fill=black!20]|\\
|[fill=black!20]| & \ & |[fill=black!20]| & |[fill=black!20]| & |[fill=black!20]|\\
\ & \ & \ & \ & \ \\
|[fill=black!20]| & \ & |[fill=black!20]| & |[fill=black!20]| & |[fill=black!20]|\\
|[fill=black!20]| & \ & |[fill=black!20]| & |[fill=black!20]| & |[fill=black!20]|\\
};

\draw[dashed,gray] (S-5-3.south east) -- (S-1-3.north east);
\draw[dashed,gray] (S-5-1.north west) -- (S-5-5.north east);
\draw[] (S-1-2.north west) -- (S-5-2.south west);
\draw[] (S-1-2.north east) -- (S-5-2.south east);
\draw[] (S-3-1.north west) -- (S-3-5.north east);
\draw[] (S-3-1.south west) -- (S-3-5.south east);

\draw[draw=black, ->] (-1.5,0.3) -- (-2.5,0.3);
\draw[draw=black, ->] (-0.5,0.3) -- (2.5,0.3);
\draw[draw=black, dashed] (-0.5,0.3) -- (-1.5,0.3);

\draw[draw=black, ->] (-1.3,0.5) -- (-1.3,2.5);
\draw[draw=black, ->] (-1.3,-0.5) -- (-1.3,-2.5);
\draw[draw=black, dashed] (-1.3,-0.5) -- (-1.3,0.5);

\draw[decorate,decoration={brace,amplitude=10pt}] (S-1-1.north west) -- (S-1-4.north west);
\draw[decorate,decoration={brace,amplitude=10pt},rotate=90] (S-1-5.north east) -- (S-5-5.north east);
\node[above,yshift=2ex] (col) at (S-1-2.north) {$(s+1)$ level-0 columns};
\node[rotate=270,above,yshift=2ex] (row) at (S-2-5.south east) {$(r+1)$ level-0 rows};
\node at (1,0.05) {$Row$};
\node[rotate=90] at (-1.05,1.4) {$Col$};
\node[] at (-0.9,-0.1) {$\neq 0$};
\node[xshift=-5ex] (M=) at (S-3-1.west) {$\widetilde{M}=$};
\end{tikzpicture}

\end{minipage}
  
  \caption{Visualisation of the matrices $M$ and $\widetilde{M}$}
  \label{fig:M}
\end{figure}

With the submatrix $M$ given, we construct a $(d+2)\times(d+2)$-submatrix $\widetilde{M}$ of $\mathcal{B}$ as
follows:
\begin{itemize}
\item Removing $(2n - d - 2)$ rows and columns from $\mathcal{B}$ results in $\widetilde{M}$.
\item There is a level-0 row (which we call $Row$) and a level-0 column (called $Col$) in $\widetilde{M}$, such that
  removing $Row$ and $Col$ in $\widetilde{M}$ yields $M$. Hence, $\widetilde{M}$ contains $(r+1)$ level-0 rows and
  $(s+1)$ level-0 columns.
\item The construction of $\widetilde{M}$ might not be unique, but is always possible. This can be seen as
  follows: First, there is at least one level-0 row of $\mathcal{B}$ that
  is not part of $M$ (otherwise, $M$ would contain all level-0 rows of $\mathcal{B}$ which means that
  $\det M \in \mathcal{I}_D$ which contradicts our assumption). Furthermore, there are at least two level-0
  columns of $\mathcal{B}$ that are not part of $M$, because $\bar s \geq 2$ and hence,
  $s \leq d - 1$.\footnote{$\mathcal{B}$ has more than $d$ level-0 columns, since $n>d$.} In particular, we can
  always choose $\widetilde{M}$ such that the intersection of $Row$ and $Col$ contains a non-zero Mandelstam
  variable.
\end{itemize}
The construction of the matrix $\widetilde{M}$ is visualised in Figure~\ref{fig:M}.

For $\widetilde{M}$, Cramers rule states that
\begin{equation}
  \label{eq:cramer}
  \mathbb{I}_{(d+2)\times (d+2)} \det \widetilde{M} - \widetilde{M} \cdot C^T = 0,
\end{equation}
where $C = (c_{ij})$ denotes the cofactor matrix of $\widetilde{M} = (\widetilde{m}_{ij})$. In particular,
$c_{ij}$ is (up to a factor of $\pm 1$) equal to the determinant of the $(d+1)\times(d+1)$-submatrix obtained by
deleting the $i$-th row and the $j$-th column from $\widetilde{M}$. In other words, $c_{ij}$ is a
$(d+1)\times(d+1)$-minor of $\mathcal{B}$, hence $c_{ij} \in \mathcal{I}_S$. In the following, we consider only
part of Eq.~(\ref{eq:cramer}):
\begin{equation}
  \label{eq:cramer-part}
  \delta_{ji} \det \widetilde{M} - \sum_{k = 1}^{s+1} \widetilde{m}_{jk} c_{ik} - \sum_{k = s+2}^{d+2}
  \widetilde{m}_{jk} c_{ik} = 0 \, \qquad i = 1,\ldots,s+1,\,j \in J\,.
\end{equation}
Here, $J$ is a (non-unique) subset of $s+1$ level-$0$ rows that contains $Row$. In other words,
\begin{equation*}
  J \subset \{1,\ldots, r+1 \}\,,\qquad |J| = s+1\,,\qquad Row \in J\,.
\end{equation*}

Performing a Laplace expansion of $\det \widetilde{M}$ along the last column of $\widetilde{M}$ (which is of level~1
because of $\bar s \geq 2$), we deduce that $\det \widetilde{M}$ is a linear combination of Schouten identities of
level $\bar r + \bar s - 1$ and $\bar r + \bar s - 2$. Hence,
\begin{equation*}
  \det \widetilde{M} \in I(\bar r + \bar s - 1) + I(\bar r + \bar s - 2)\,.
\end{equation*}
Furthermore, in the third term of Eq.~(\ref{eq:cramer-part}), the Schouten identities $c_{ik}$ with $k > s+1$
are of level $(\bar r + \bar s - 1)$. We therefore conclude that the middle term is an element in the
following ideal:
\begin{equation}
  \label{eq:sum=in-I}
  \text{for all}\  i = 1, \ldots ,s+1,\,j \in J \, :\quad  \left(\sum_{k = 1}^{s+1} \widetilde{m}_{jk} c_{ik} \right) \in I(\bar r + \bar s - 1)
  + I(\bar r + \bar s - 2)\,. 
\end{equation}

Now, denote by $N = (\widetilde{m}_{jk})$ (with $j \in J$ and $k \in \{1, \ldots ,s+1\}$) the
$(s+1)\times(s+1)$-submatrix of $\widetilde{M}$ that occurs in Eq.~(\ref{eq:sum=in-I}). It is also a submatrix
of $\mathcal{S}$ because it only consists of level-0 rows and columns. Since $s + 1 \leq d$, we deduce that
$\det N \in \Mi(\mathcal{S})$.\footnote{In the case that $s = 0$, $N$ is just a Mandelstam variable. But within
  the construction of $\widetilde{M}$, we chose $Row$ and $Col$ such that its intersection (which is $N$ in that
  case) is non-zero. Therefore, $\det N = N \neq 0$.} In particular, $\det N \neq 0$ and by inverting $N$ in
Eq.~(\ref{eq:sum=in-I}) using Cramers rule, we find
\begin{equation*}
  \text{for all}\  j \in J,\, k \in \{1, \ldots ,s+1\} \, :\quad \det N \cdot c_{jk} \in I(\bar r + \bar s - 1) + I(\bar r + \bar s - 2)\,.
\end{equation*}

Finally, setting $j = Row$ and $k = Col$, we have $c_{jk} = \det M$ -- which corresponds to the Schouten
identity of level $(\bar r + \bar s)$ we started with. It follows directly that for all $\det M \in I(\bar r +
\bar s)$, either $\det M \in \mathcal{I}_D$ or
\begin{equation}
  \label{eq:recursiveMAIN}
  \exists   \det N \in \Mi(\mathcal{S}) \, : \quad \det N \cdot
  \det M \in 
  I(\bar r + \bar s - 1) + I(\bar r + \bar s - 2) \,.
\end{equation}
In other words,
\begin{equation*}
  \exists \, \Delta \in \Mon[\Mi(\mathcal{S})] \, : \quad \Delta \cdot I(\bar r + \bar s) \subset I(\bar r + \bar s -
  1) + I(\bar r + \bar s - 2) + \mathcal{I}_D
\end{equation*}
and a recursion over $\bar r$ and $\bar s$ proves the general statement in Eq.~(\ref{eq:MAIN}).

\subsubsection{Proof of Eq.~(\ref{eq:I-Delta})}
\label{sec:schouten-ideal-2}

We prove Eq.~(\ref{eq:I-Delta}) in three steps. It directly follows from
Eq.~(\ref{eq:MAIN}), as well as Eqs.~(\ref{eq:PartA}, \ref{eq:PartB} and \ref{eq:PartC}).

\paragraph*{Part 1:}

First of all, we show that
\begin{equation}
  \label{eq:PartA}
  \exists \, \Delta \in \Mon[\Mi(\mathcal{S})] \, : \quad \Delta I(2) \subset I(0) + I(1) + \left
    \langle \det B_2(i,j) \,\big|\, i,j = 1,\ldots , n \right \rangle\,.
\end{equation}
For this purpose, consider an arbitrary level-2 Schouten identity $\det M \approx 0$, with $M$ being a
$(d+1)\times(d+1)$-submatrix of $\mathcal{B}$. As explained in the previous section, we may assume that $M$ has
either one ($\bar r = 1$, $\bar s = 1$) or two level-1 columns ($\bar r = 0$, $\bar s = 2$). In the latter case,
the proof of the previous section goes through and Eq.~(\ref{eq:recursiveMAIN}) is satisfied. Hence, we only
need to consider the other case $(\bar r = \bar s = 1)$. In particular, we show that if $M$ contains the level-1
row $i+n$ and the level-1 column $j+n$ of $\mathcal{B}$, then:
\begin{equation}
  \label{eq:proposition1}
  \exists \, \Delta \in \Mon[\Mi(\mathcal{S})] \,:\quad \Delta \det M \in I(1) + \langle
  \det B_2(i,j) \rangle\,,
\end{equation}
which implies Eq.~(\ref{eq:PartA}).

We prove Eq.~(\ref{eq:proposition1}) by induction. For the following discussion, we fix $i$ and $j$. Let $I_2(K,L)$ be the ideal
generated by all level-2 Schouten identities $\det M \approx 0$, such that the $(d+1)\times(d+1)$-submatrix $M$
of $\mathcal{B}$ has the following properties:
\begin{itemize}
\item[i)] $M$ contains the level-1 row $i+n$ and the level-1 column $j+n$ of $\mathcal{B}$.
\item[ii)] $K$ rows ($L$ columns) of $M$ stem from the rows $i,\ldots, i + d - 1$ (columns
  $j,\ldots, j + d - 1$) (modulo $n$) of $\mathcal{B}$.
\end{itemize}
Hence, $I_2(d,d) = \langle \det B_2(i,j) \rangle$ and Eq.~(\ref{eq:proposition1}) is a recursive consequence of
the following two propositions:
\begin{align}
  \label{eq:ind1}
  \text{if } K < d\,,\quad \exists \, \Delta \in \Mon[\Mi(\mathcal{S})] \,:&\quad \Delta \cdot I_2(K,L) \subset I(1)
  + I_2(K+1,L)\,,\\
  \label{eq:ind2}
  \text{if } L < d\,,\quad\exists\,  \Delta \in \Mon[\Mi(\mathcal{S})] \,:&\quad \Delta \cdot I_2(K,L) \subset I(1) +
  I_2(K,L+1)\,.
\end{align}

Here, we present the proof of Eq.~(\ref{eq:ind2}). Eq.~(\ref{eq:ind1}) can be shown in the same way, except that
the roles of rows and columns are interchanged. We start with a $(d+1)\times(d+1)$-submatrix $M$ of
$\mathcal{B}$, such that $\det M \in I_2(K,L)$ with $L<d$. For $M$ given, we construct a
$(d+2)\times(d+2)$-matrix $\widetilde{M}$ as follows:
\begin{itemize}
\item Removing the first row from $\widetilde{M}$ yields a $(d+1)\times(d+2)$-submatrix $\widehat{M}$ of
  $\mathcal{B}$.
\item There is a unique $k_0 \in \{1,\ldots , d+1\}$, such that removing the $k_0$-th column from $\widehat{M}$
  yields $M$. We construct $\widetilde{M}$ such that the $k_0$-th column stems from one of the columns
  $j,\ldots, j + d - 1$ (modulo $n$) of $\mathcal{B}$, which is possible because $L < d$.
\item Note that $\widehat{M}$ has $d+1$ level-0 columns. Hence, at least one of those cannot stem from one of
  the columns $j,\ldots, j + d - 1$ (modulo $n$) of $\mathcal{B}$. Let us agree that at least the $l_0$-th
  column has this property. Obviously, $l_0 \neq k_0$.
\item The first two rows of $\widetilde{M}$ coincide, hence, $\det \widetilde{M} = 0$.
\end{itemize}

Now, Cramers rule states that $\widetilde{M}C^T = 0$, where $C = (c_{kl})$ is the cofactor matrix of
$\widetilde{M} = (\widetilde{m}_{kl})$. The first column of this matrix equation reads
\begin{equation*}
  \sum_{l = 1}^{d+2} \widetilde{m}_{kl} c_{1l} = 0\,.
\end{equation*}
Note that up to a factor of $\pm 1$, $c_{1l}$ is the determinant of the $(d+1)\times(d+1)$-matrix obtained by
removing the first row and the $l$th column from $\widetilde{M}$. Hence, $c_{1l} \in I(2)$ for $l \leq d+1$ and
$c_{1\, d+2} \in I(1)$. In particular, $c_{1k_0} \propto \det M \in I_2(K,L)$ and
$c_{1l_0} \propto \det M_{l_0 \to k_0} \in I_2(K,L+1)$, where the matrix $M_{l_0 \to k_0}$ differs from $M$ by
only one column. Indeed, it contains the $k_0$-th column of $\widetilde{M}$ instead of the $l_0$-th. We deduce
that
\begin{equation*}
  \sum_{1 \leq l \leq d+1 }^{l \neq l_0} \widetilde{m}_{kl} c_{1l} \in I(1) + I_2(K,L+1)\,.
\end{equation*}

Now, consider only the rows $2,\ldots, d+1$ of that relation. The matrix $N = (\widetilde{m}_{kl})$ with
$k \in \{ 2,\ldots, d+1 \}$ and $l \in \{1,\ldots , d+1\} \backslash \{l_0\}$ is a $(d\times d)$-submatrix of
$\mathcal{S}$ and can hence, be inverted using Cramers rule. We find that
\begin{equation*}
  \det N \cdot c_{1l}  \in I(1) + I_2(K,L+1)
\end{equation*}
and setting $l = k_0$ finally proves Eq.~(\ref{eq:ind2}).

\paragraph{Part 2:}

In a second step, we show that
\begin{equation}
  \label{eq:PartB}
  \exists\,  \Delta \in \Mon[\Mi(\mathcal{S})] \quad : \quad \Delta I(1) \subset I(0) + \left \langle \det B_1(i,j)
    \,|\, i,j = 1,\ldots , n\right \rangle\,.
\end{equation}
The proof is similar to the previous one. Let $\det M \approx 0$ be an arbitrary level-1 Schouten identity,
where $M$ is a $(d+1)\times(d+1)$-submatrix of $\mathcal{B}$. As explained in the previous section, we may
assume that $M$ has exactly one level-1 column ($\bar r = 0, \bar s = 1$). In particular, we show that if $M$ contains the
level-1 column $i+n$ of $\mathcal{B}$, then:
\begin{equation}
  \label{eq:proposition2}
  \exists\,  \Delta \in \Mon[\Mi(\mathcal{S})] \,:\quad \Delta \det M \in I(0) + \langle
  \det B_1(i,j) \,|\, j = 1,\ldots , n \rangle\,,
\end{equation}
which implies Eq.~(\ref{eq:PartB}).

Again, we prove Eq.~(\ref{eq:proposition2}) by induction for a fixed $i$. Let $I_1(K,L)$ be the ideal
generated by all level-1 Schouten identities $\det M \approx 0$, such that the $(d+1)\times(d+1)$-submatrix $M$
of $\mathcal{B}$ has the following properties:
\begin{itemize}
\item[i)] $M$ contains the level-1 column $i+n$ of $\mathcal{B}$.
\item[ii)] $K$ rows ($L$ columns) of $M$ stem from the rows (columns) $i,\ldots, i + d - 1$ (modulo $n$) of
  $\mathcal{B}$.
\end{itemize}
Hence, $I_{1}(d,d) = \langle \det B_1(i,j) \,|\, j = 1,\ldots , n \rangle$ and Eq.~(\ref{eq:proposition2}) is a
recursive consequence of the following two propositions:
\begin{align}
  \label{eq:ind3}
  \text{if } K < d\,,\quad\exists\,  \Delta \in \Mon[\Mi(\mathcal{S})] \,:&\quad \Delta \cdot I_1(K,L) \subset I(0)
  + I_1(K+1,L)\,,\\
  \label{eq:ind4}
  \text{if } L < d\,,\quad\exists\,  \Delta \in \Mon[\Mi(\mathcal{S})] \,:&\quad \Delta \cdot I_1(K,L) \subset I(0) +
  I_1(K,L+1)\,.
\end{align}

Here, we give the proof of Eq.~(\ref{eq:ind4}), Eq.~(\ref{eq:ind3}) follows analogously. For a given
$(d+1)\times(d+1)$-submatrix $M$ of $\mathcal{B}$, such that $\det M \in I_1(K,L)$ with $L<d$, we construct a
$(d+2)\times(d+2)$-matrix $\widetilde{M}$ as follows:
\begin{itemize}
\item Removing the first row from $\widetilde{M}$ yields a $(d+1)\times(d+2)$-submatrix $\widehat{M}$ of
  $\mathcal{B}$.
\item There is a unique $k_0 \in \{1,\ldots , d+1\}$, such that removing the $k_0$-th column from $\widehat{M}$
  yields $M$. Again, we construct $\widetilde{M}$ such that the $k_0$-th column stems from one of the columns
  $i,\ldots, i + d - 1$ (modulo $n$) of $\mathcal{B}$, which is possible because $L < d$.
\item Note that $\widehat{M}$ has $d+1$ level-0 columns. Hence, at least one of those (say, the $l_0$-th) cannot
  stem from one of the columns $i,\ldots, i + d - 1$ (modulo $n$) of $\mathcal{B}$. Again, $l_0 \neq k_0$.
\item The first two rows of $\widetilde{M}$ coincide, hence, $\det \widetilde{M} = 0$.
\end{itemize}

Now, Cramers rule states that $\widetilde{M}C^T = 0$, where $C = (c_{kl})$ is the cofactor matrix of
$\widetilde{M} = (\widetilde{m}_{kl})$. Considering the first column of this matrix equation, we have
\begin{equation*}
  \sum_{l = 1}^{d+2} \widetilde{m}_{kl} c_{1l} = 0\,.
\end{equation*}
Up to a factor of $\pm 1$, $c_{1l}$ is the determinant of the $(d+1)\times(d+1)$-matrix obtained by removing the
first row and the $l$th column from $\widetilde{M}$. Hence, $c_{1l} \in I(1)$ for $l \leq d+1$ and
$c_{1\, d+2} \in I(0)$. In particular, $c_{1k_0} \propto \det M \in I_1(K,L)$ and
$c_{1l_0} \propto \det M_{l_0 \to k_0} \in I_1(K,L+1)$, where the matrix $M_{l_0 \to k_0}$ differs from $M$ by
only one column (it contains the $k_0$-th column of $\widetilde{M}$ instead of the $l_0$-th). We deduce
that
\begin{equation*}
  \sum_{1\leq l \leq d+1}^{l \neq l_0} \widetilde{m}_{kl} c_{1l} \in I(0) + I_1(K,L+1)\,.
\end{equation*}

Again, we only consider the rows $2,\ldots, d+1$ of that relation. The matrix $N = (\widetilde{m}_{kl})$ with
$k \in \{ 2,\ldots, d+1 \}$ and $l \in \{1,\ldots , d+1\} \backslash \{l_0\}$ is a $(d\times d)$-submatrix of
$\mathcal{S}$ and can be inverted using Cramers rule. Finally,
\begin{equation*}
\det N \cdot c_{1l}  \in I(0) + I_1(K,L+1)
\end{equation*}
and setting $l = k_0$ proves Eq.~(\ref{eq:ind4}).

\paragraph{Part 3:}

Finally, we prove that for any $i \in \{1,\ldots , n\}$,
\begin{equation}
  \label{eq:PartC}
  \det N_{ii} \sum_{j = 1}^n y_{ij} \in \left \langle \sum_{j = 1}^n s_{ij} \right
  \rangle + \left \langle \det B_1(i,j) \right
  \rangle\,,
\end{equation}
where $N_{ii} \in \Mi(\mathcal{S})$ is defined in Eq.~(\ref{eq:N_ij}).

Fix $i \in \{1,\ldots , n\}$. Then, for any $j \in \{1,\ldots , n\}$, let $N_{ii}(k \to j)$ be the matrix
$N_{ii}$, where the $(k+1)$st row is replaced by $
\begin{pmatrix}
  s_{ji} & s_{j i + 1} & \cdots & s_{j i + d - 1}  
\end{pmatrix}
$. In particular,
\begin{equation}
  \label{eq:propertyN}
  \sum_{j = 1}^n \det N_{ii} ( k \to j ) \in \left \langle \sum_{j = 1}^n s_{ij} \right
  \rangle\,,
\end{equation}
because the determinant of $N_{ii} ( k \to j )$ is linear (especially in the $(k+1)$st row).

Now, a Laplace expansion of $ \det B_1(i,j)$ (see~Eq.~(\ref{eq:B1})) with respect to the last column results in
\begin{equation*}
  \det B_1(i,j) =  y_{ij} \det N_{ii} - \sum_{k = 0}^{d-1} y_{ik+1} \det N_{ii} ( k \to j )\,,
\end{equation*}
which holds for all $j \in \{1 ,\ldots , n\}$. In particular,
\begin{equation*}
  \det N_{ii} \sum_{j = 1}^n y_{ij} = \sum_{j = 1}^n \det B_1(i,j) + \sum_{k = 0}^{d-1} y_{ik+1} \sum_{j = 1}^n
  \det N_{ii} ( k \to j )\,,
\end{equation*}
which, taking Eq.~(\ref{eq:propertyN}) into account, proves Eq.~(\ref{eq:PartC}).

\section{Parity-Odd Vertices}
\label{sec:parity-odd-vertices}

So far, we only discussed parity-even vertices, i.e.\ terms in the Lagrangian which do not involve the epsilon
tensor $\epsilon_{\mu_1 \cdots \mu_d}$. However, the discussion of the previous sections can simply be
generalised also for parity-odd vertices. 

First of all, the most general form of a parity-odd vertex is given by Eq.~(\ref{eq:L^n}) but with $\mathcal{V}$
replaced by
\begin{equation}
  \label{eq:Vtilde}
  \tilde{\mathcal{V}} = \sum_{I_1 \cdots I_d} Q_{I_1\cdots I_d} \tilde{\mathcal{V}}^{I_1\cdots I_d}\,,
\end{equation}
where $\tilde{\mathcal{V}}^{I_1 \cdots I_d} \in \mathbb{R}[y_{ij}, z_{ij}|_{i \leq j}, s_{ij} |_{i \leq j}]$
contains the parity-even contractions\footnote{We discussed these in the previous sections where they were
  called $\mathcal{V}$.} and
\begin{equation}
  \label{eq:Q-parity-odd}
  Q_{I_1 \cdots I_d} = \epsilon_{\mu_1\cdots \mu_d} b^{\mu_1}_{I_1} \cdots b^{\mu_d}_{I_d}
\end{equation}
is totally antisymmetric in its indices ($I_k \in\{ 1,\ldots ,2n\}$). The derivative operators $b_I$ were introduced
in Section~\ref{sec:equiv-relat-vert}, right before Eq.~(\ref{defofB}). Note that for $i = 1,\ldots , n$, we
have $b_i = P_i$ and $b_{i+n} = A_i$. The structure of the gauge-invariant parity-odd vertices depends on the dimension:
\begin{itemize}
\item For $d\geq 2n$, there are no parity-odd $n$-point vertex operators, because $Q_{I_1 \cdots I_d} = 0$
(the vector $b$ has only $2n-1$ independent entries up to total derivatives).

\item In the case $n > d$, we again make use of the fact that we consider $[\tilde{\mathcal{V}}]$ in the ring of
  fractions. The crucial point is that the general form of an elementary building block $Q_{I_{1}\dotsb I_{d}}$
  of parity-odd vertices can be highly simplified, when it is multiplied with the upper-left $d\times d$
  submatrix of $\mathcal{S}$. Denote this matrix by $S_d$. Its determinant,
\begin{equation*}
  \det S_d = \frac{1}{d!} \epsilon_{\mu_1 \cdots \mu_d} \epsilon_{\nu_1 \cdots \nu_d} b_1^{\mu_1} \cdots
  b_d^{\mu_d} b_1^{\nu_1} \cdots b_d^{\nu_d}\,,
\end{equation*}
is a non-zero minor of $\mathcal{S}$, hence, $\det S_d \in \Mi(\mathcal{S})$ and we conclude that
\begin{align*}
  \det S_d \cdot Q_{I_1 \cdots I_d} &=
  \left(
    \mathcal{B}_{1I_1} \cdots \mathcal{B}_{d I_d}
  \right)\big|_{[I_1 \cdots I_d]} \cdot Q_{1\cdots d}\,.
\end{align*}
In other words, for any parity-odd vertex in the Lagrangian given by the vertex generating operator
$\tilde{\mathcal{V}}$ in Eq.~(\ref{eq:Vtilde}), we find
\begin{equation}
  \label{eq:Q123}
  \det S_d \cdot \tilde{\mathcal{V}} = Q_{1 \cdots d} \cdot \mathcal{V}\,,
\end{equation}
where $\mathcal{V} \in \mathbb{R}[y_{ij}, z_{ij}|_{i \leq j}, s_{ij} |_{i \leq j}]$ as in the parity-even case.

Now, since we work in the ring of fractions, we can divide by $\det S_d \in \Mi(\mathcal{S})$. Furthermore, $Q_{1 \cdots d}$ is gauge invariant:
\begin{equation*}
  [Q_{1\cdots d},a_k \cdot P_k] = 0\,.
\end{equation*}
Hence, along the same lines as in Section~\ref{sec:case-n-}, we find that
\begin{equation}
  \label{eq:PY-general-odd}
  \tilde{\mathcal{V}} \approx Q_{1\cdots d} \cdot \mathcal{P}_\mathcal{V} (Y_i^j , s_{ij})\,.
\end{equation}

\item The intermediate case, $n \leq d \leq 2n - 1$, can be tackled in a similar way. Let $S_{n-1}$ be the upper
  left $(n-1)\times(n-1)$ submatrix of $\mathcal{S}$. Up to a factor, its determinant is given by
  \begin{equation*}
    \det S_{n-1} \,\,\propto\,\, \epsilon_{\mu_1\cdots \mu_d}\epsilon_{\nu_1\cdots \nu_{n-1}}{}^{\mu_n \cdots \mu_d} b_1^{\mu_1} \cdots
    b_{n-1}^{\mu_{n-1}} b_1^{\nu_1} \cdots b_{n-1}^{\nu_{n-1}}\,.
  \end{equation*}
  Multiplying it to the general vertex in Eq.~(\ref{eq:Vtilde}) yields
  \begin{multline*}
    \det S_{n-1} \tilde{\mathcal{V}} \,\,
\propto\,\, \sum_{I_1 \cdots I_d} \epsilon_{\nu_1 \cdots \nu_{n-1}}{}^{\mu_n \cdots
      \mu_d} \,b_1^{\mu_1} \cdots b_{n-1}^{\mu_{n-1}}\, b_1^{\nu_1} \cdots b_{n-1}^{\nu_{n-1}}\, b_{I_1\,\mu_{1}}
     \cdots  b_{I_d\,\mu_{d}} \,\tilde{\mathcal{V}}^{[I_1\cdots I_d]}\,.
  \end{multline*}
  We again work in the ring of fractions. Hence, we can divide by $\det S_{n-1}$, because it is a nonzero
  minor of $\mathcal{S}$. We finally find
  \begin{equation*}
    \tilde{\mathcal{V}} \approx \sum_{I_n \cdots I_d} Q_{1\cdots n-1 \, I_n \cdots I_d} \hat{\mathcal{V}}^{I_n
      \cdots I_d}\,.
  \end{equation*}
  Here, we collect all direct parity even index contractions into one vertex generating operator
  $\hat{\mathcal{V}}^{I_n \cdots I_d}$, which is fully antisymmetric in its indices. Note that
  $I_n,\ldots,I_d > n$ because the $Q$-tensor is fully antisymmetric.\footnote{If one of those indices equals
    $n$, $Q_{1\cdots n \,I_{n+1} \cdots I_d}$ vanishes equivalently because it equals a total
    derivative.} In particular, with Eq.~(\ref{eq:Q-parity-odd}), the $Q$-tensor reduces to a ``square-root of
  a Horndeski-type operator'',
  \begin{equation*}
    \label{SHO}
    Q_{1\cdots n-1 \, I_n \cdots I_d} = \epsilon_{\mu_1\cdots \mu_d} P^{\mu_1}_{1} \cdots P^{\mu_{n-1}}_{n-1}
    A^{\mu_n}_{I_n - n} \cdots A^{\mu_d}_{I_d - n}\,.
  \end{equation*}
  It is trivially gauge invariant up to total derivatives:
  \begin{equation*}
    [Q_{1\cdots n-1 \, I_n \cdots I_d},a_k \cdot P_k] \approx 0\,.
  \end{equation*}

Hence, as in the case $n>d$, we conclude that 
  \begin{equation*}
    \tilde{\mathcal{V}} \approx \sum_{I_n \cdots I_d} Q_{1\cdots n-1 \, I_n \cdots I_d} \cdot \mathcal{P}_\mathcal{V}^{I_n \cdots I_d} (Y_i^j , s_{ij})\,,
  \end{equation*}
  where the polynomials $\mathcal{P}_\mathcal{V}^{I_n \cdots I_d}$ only depend on $Y_i^j$ and the Mandelstam variables. E.g. for
  $d=2n-1$, there is only one term in the sum, namely
  \begin{equation*}
    \tilde{\mathcal{V}} \approx Q_{1\cdots n-1\,n+1 \cdots 2n} \cdot \mathcal{P}_\mathcal{V}^{n+1 \cdots 2n} (Y_i^j , s_{ij})\,.
  \end{equation*}
  Note that $Q_{1\cdots n-1\,n+1 \cdots 2n}$ squares to the Lovelock operator \eqref{Lovelock}. This covers also
  the case of $n=3$ and $d=5$, where $\mathcal{P}_\mathcal{V}$ is parity-even cubic vertex operator \cite{Manvelyan:2010jr}. These covariant parity-odd $5d$ vertices match the light-cone classification \cite{Metsaev:2005ar}.
\end{itemize}

In all cases the vertices can be brought to a form in which they are gauge invariant without the use of Schouten identities. We conclude, that the situation with parity-odd vertices is analogous to the parity-even ones: Schouten identities do not give rise to new vertices except for the cubic ones in three dimensions, studied in \cite{Kessel:2018ugi}.

\section{Discussion}
\label{sec:discussion}

In this work, we complete the classification of independent vertices of arbitrary order $n\geq 3$ for massless bosonic fields with arbitrary spin in arbitrary spacetime dimensions  $d\geq 2$.\footnote{We concentrate on the traceless-transverse (TT) part of the vertices for classification, as discussed in the beginning of Section \ref{sec:preliminaries}.}

We briefly summarise the results:
\begin{itemize}
\item For dimensions $d\geq 2n-1$ there are no non-trivial Schouten identities. After reducing to the independent Mandelstam variables, we find that all gauge invariant operators can be expressed as polynomials in the gauge-invariant combinations $c_{ij}$ and $Y_{i}^{j}$,
\begin{equation}
\mathcal{V}  \in M_{1}^{-1}\mathbb{R}[s_{ij},c_{ij},Y_{i}^{j}]\, ,
\end{equation}
where $M_{1}$ is the set of all products of Mandelstam variables $s_{ij}$ ($i\not = j$). The invariant combinations $Y_{i}^{j}$ are labelled by $i=1,\dotsc ,n$ and $j=2,\dotsc ,n-2$. 
\item For dimensions $d < n$ we have the full set of Schouten identities at our disposal. All gauge invariant operators are already generated by the $Y_{i}^{j}$'s, where $i=1,\dotsc ,n$ and $j=2,\dotsc ,d-1$. All remaining relations are generated by level-$0$ Schouten identities and specific quadratic expressions $q_{2}^{i}$ in the variables $Y_{i}^{j}$,
\begin{equation}
[\mathcal{V}] \in \frac{M^{-1}\mathbb{R}[s_{ij},Y_{i}^{j}]}{\langle (\det B_{0} (A)), q_{2}^{i} \rangle} \, ,
\end{equation}
where again we reduced to the independent Mandelstam variables.
\item In the intermediate case ($2n - 1 > d \geq n$), we have Schouten identities, but because $d\geq n$ the
  non-trivial Schouten identities involve at least $(d-n)+2\geq 2$ level-$1$ rows and columns. By an argument
  analogous to the one leading to Eq.~\eqref{eq:MAIN} one can show that in the ring of fractions all Schouten
  identities are generated by those that contain $n-1$ level-$0$ rows and columns and $(d-n)+2$ rows and columns
  of level-$1$. Let us denote them by $\det B_{2 (d-n)+4} (A)$, where $A$ labels the possible choices of the
  level-$1$ rows and columns. These generators are all gauge-invariant (up to total derivatives), and hence we
  can express them in terms of the invariant combinations $c_{ij}$ and $Y_{i}^{j}$ as in
  Section~\ref{sec:case-2n-leq}. Then the gauge invariant vertices are classified by equivalence classes
\begin{equation}
[\mathcal{V}] \in \frac{M^{-1}\mathbb{R}[s_{ij},c_{ij},Y_{i}^{j}]}{\langle \det B_{2 (d-n)+4} (A)\rangle}\, .
\end{equation} 
\end{itemize}

An interesting question is whether the higher order vertices can induce deformations of gauge transformations
for the fields involved. Deformations arise when the gauge variation is non-trivial before imposing the
equations of motion. Terms in the variation that contain the equations of motion have to be compensated by a
non-trivial $\delta^{(n-2)}$ in~Eq.~\eqref{NE}. We have found that in all dimensions, as long as we are allowed
to divide by Mandelstam variables, the independent gauge-invariant vertices can be expressed in terms of the
combinations $c_{ij}$ and $Y_{i}^{j}=c_{i,i+j i+1}$, but these -- as defined in Eq.~\eqref{defc2} and
Eq.~\eqref{defc3} -- are manifestly gauge-invariant without need of the equations of motion. This strongly
suggests that the vertex does not induce a deformation. Strictly speaking we can only conclude that
$\Delta \mathcal{V}$ for an appropriate product $\Delta $ of Mandelstam variables does not induce any
deformation. However, in Fourier space $\Delta$ is simply a (generically non-zero) number and should not change
the general structure of deformations, hence we do not expect that $\mathcal{V}$ itself can induce a
deformation.\footnote{This fits nicely with the observation obtained within the BRST formalism for spins up to $s=4$ that deformations of the gauge algebra can only arise from cubic vertices~\cite{Boulanger:2000rq,Bekaert:2005jf,Boulanger:2008tg} and observations about some quartic vertices~\cite{Taronna:2017wbx,Roiban:2017iqg}.}

To recapitulate, as soon as we allow for dividing by Mandelstam variables (and hence, we loose manifest
locality), the independent vertices of order $n\geq 4$ can be all written in terms of linearised curvatures of
HS fields. Therefore they are manifestly gauge invariant with respect to linearised gauge transformations and do
not introduce deformations for the latter. On the other hand, if such deformations of the gauge transformations,
induced from cubic vertices, exist in the theory, then these vertices will be completed by further non-linear
terms. This is similar to higher-curvature terms in Einstein Gravity, whose non-linear structure is gauge
invariant with respect to full diffeomorphisms, induced from the Einstein-Hilbert cubic vertex. Such non-linear
completions may make use of a non-linear generalisation of de Wit-Freedman curvatures \cite{deWit:1979sib},
which are not known in the metric-like formulation (see, however, \cite{Manvelyan:2010jf}). In the frame
formulation, these vertices would correspond to structures that make use of Weyl tensors and their descendants
(zero form sector of the Vasiliev system). In light of our findings here, the three dimensional results of
\cite{Fredenhagen:2019hvb} can be interpreted as a particular case of the general dimensional results: all the
independent vertices are given through linearised curvatures, which are on-shell trivial in $d=3$. \smallskip

Even though the classification is done for Minkowski spaces, we expect the vertices found here to deform
smoothly to $(A)dS$ spacetimes as it happens for cubic vertices. Indeed, the existence of $(A)dS$ extensions
for linearised de Wit-Freedman curvatures for HS fields \cite{Manvelyan:2007hv} allows to straightforwardly lift
vertices given through curvatures to $(A)dS_d$. This is also true for the operators \eqref{Lovelock} and their parity-odd counterparts given in Section 
\ref{SHO}, where one can simply replace derivatives with $(A)dS_d$ covariant ones. \smallskip

Our results should have a direct analogue for correlation functions of conserved tensors in $d-1$ dimensional conformal field theories, which can be classified with similar methods~\cite{Costa:2011mg}. For $n=3$ there is a precise match between independent vertices and three-point functions~\cite{Metsaev:2005ar,Manvelyan:2010jr,Giombi:2011rz,Costa:2011mg,Joung:2011ww,Conde:2016izb,Francia:2016weg,Sleight:2017fpc,Fredenhagen:2018guf}. It would be interesting to compare our findings for $n\geq 4$ with the group theoretic results of~\cite{Kravchuk:2016qvl}.

Next, we would like to note that there is another interpretation of Eq.~(\ref{eq:gauge-0}) which we solved here.
One can think of Eq.~(\ref{eq:gauge-0}) as a Ward identity for an $n$-point amplitude computed in a theory of
interacting HS fields. It is clear from our discussion, that the building blocks of the amplitudes are given
through $c_{ij}$, $Y_{i}^{j}=c_{i,i+j i+1}$ and Mandelstam variables, including negative powers of the latter.
They correspond to arbitrary tensor contractions of linearised curvatures \cite{deWit:1979sib} of HS gauge
fields and their derivatives. These linear de Wit-Freedman curvatures (or their traceless part: the Weyl
tensors) and their derivatives are the only on-shell non-zero gauge invariants with respect to the linearised
gauge transformations. It is natural that the amplitudes for $n\geq 4$ should be given through gauge invariant
quantities, as they are observable.

The amplitude interpretation might be less motivated in three dimensions since there are no propagating HS
massless particles in three dimensions. As proved in \cite{Fredenhagen:2019hvb}, there are no candidate
invariants for amplitudes with such fields either for $d=3$. There is one difference between amplitudes and
vertices though -- the latter are supposed to be local, while the former do not have to. Given that one can
always multiply the candidate invariant vertices (amplitudes) by a non-vanishing function of Mandelstam
variables, one can show that relaxing locality would not help to get non-zero amplitudes in $d=3$. There is an
interesting conclusion to be made here: since the amplitude is a sum of exchanges\footnote{The exchange is again
  a notion that is defined when there are particles to exchange, but this should not affect our argument, given
  that a propagator for massless HS fields can be formally defined in three dimensions. See, e.g.,
  \cite{David:2009xg,Giombi:2013fka}. We thank Shailesh Lal for a discussion about this point.} and contact
vertices, vanishing amplitudes imply that the exchanges and contact vertices should cancel each other. This is
only possible if the non-local parts of the exchanges sum up to zero, which should be specific to three
dimensions and is presumably due to the special structure of vertices and Schouten identities present only in
three dimensions. We plan to study the Lagrangian formulation of metric-like non-linear HS theories with(out)
matter in the near future to expose these special properties of HS gravities in $d=3$.

\paragraph{Note added} We learned from Euihun Joung and Massimo Taronna about their preprint with
related results \cite{Joung:2019wbl}, which will appear on arxiv simultaneously.

\section*{Acknowledgements}

The authors are grateful to Kostya Alkalaev, Xavier Bekaert, Dario Francia, Euihun Joung, Shailesh Lal, Ruben Manvelyan and Eugene Skvortsov for useful discussions on the subject of this work. KM is grateful to Max Planck Institute for Gravitational Physics (Albert Einstein Institute), where part of this work was done. The hospitality of the Erwin Schr{\"o}dinger International Institute for Mathematics and Physics during the program on “Higher Spins and Holography” where this work was initiated is greatly appreciated. The work of KM was supported in part by Scuola Normale, by INFN (IS GSS-Pi) and by the MIUR-PRIN contract 2017CC72MK\_003.


\end{document}